\documentclass[10pt,a4paper,twoside]{article}

\usepackage[english]{babel}
\usepackage{comment}
\usepackage[letterpaper,top=2cm,bottom=2cm,left=3cm,right=3cm,marginparwidth=1.75cm]{geometry}
\usepackage{cite}
\usepackage{xcolor}
\usepackage{hyperref}
\hypersetup{colorlinks,linktoc=page,linkcolor=purple,citecolor=blue,urlcolor=blue}
\usepackage[utf8]{inputenc}
\usepackage{amsmath}
\usepackage{amssymb}
\usepackage{bm}
\usepackage{graphicx}
\usepackage{mathtools}
\usepackage{framed}
\usepackage{wrapfig}

\renewcommand{\vec}[1]{{\rm vec}\left({#1}\right)}

\newcommand{\dd}{\text{d}}

\renewcommand{\dd}{{\rm d}}
\newcommand{\xx}{\bm{x}}
\newcommand{\uu}{\bm{u}}
\newcommand{\cc}{\bm{c}}
\newcommand{\ww}{\bm{w}}
\newcommand{\MM}{\bm{M}}
\newcommand{\RR}{\bm{R}}
\newcommand{\QQ}{\bm{Q}}
\newcommand{\mmu}{\bm{\mu}}
\newcommand{\rrho}{\bm{\rho}}

\newcommand{\vv}{\bm{v}}
\newcommand{\llambda}{\bm{\lambda}}

\newcommand{\blueflag}{0}
\newcommand{\blue}[1]{%
  \ifnum\blueflag=1
    \textcolor{blue}{#1}%
  \else
    #1%
  \fi
}

\title{\textbf{A statistical physics framework for optimal learning}}
\author{Francesca Mignacco$^{1,2,\mathsection}$, Francesco Mori$^{3,\dagger}$\\\\
  {\small
  $^{1}$Joseph Henry Laboratories of Physics, Princeton University, Princeton, NJ 08544, USA}\\
  {\small
  $^{2}$Graduate Center, City University of New York, New York, NY 10016, USA}\\
  {\small
  $^{3}$Center of Mathematical Sciences and Applications, Harvard University, Cambridge, MA 02138, USA}\\
  {\small $^\mathsection$\texttt{fmignacco@princeton.edu} $\qquad^\dagger$\texttt{francesco.mori@cmsa.fas.harvard.edu}
  }
}

\date{}

\begin{document}
\maketitle

\begin{abstract}
\noindent Learning is a complex dynamical process shaped by a range of interconnected decisions. Careful design of hyperparameter schedules for artificial neural networks or efficient allocation of cognitive resources by biological learners can dramatically affect performance. Yet, theoretical understanding of optimal learning strategies remains sparse, especially due to the intricate interplay between evolving metaparameters and nonlinear learning dynamics. The search for optimal protocols is further hindered by the high dimensionality of the learning space, often resulting in predominantly heuristic, difficult to interpret, and computationally demanding solutions. Here, we combine statistical physics with control theory in a unified theoretical framework to identify optimal learning protocols in prototypical neural network models. In the high-dimensional limit, we derive closed-form ordinary differential equations that track online stochastic gradient descent through low-dimensional order parameters. We formulate the design of learning protocols as an optimal control problem directly on the dynamics of the order parameters with the goal of minimizing the generalization error. This formulation encompasses a variety of learning scenarios, optimization constraints, and control budgets. We apply it to representative cases, including optimal curricula, adaptive dropout regularization and noise schedules in denoising autoencoders. We find nontrivial yet interpretable strategies highlighting how optimal protocols mediate learning trade-offs. Our results establish a principled foundation for understanding and designing optimal protocols and suggest a path toward a theory of meta-learning grounded in statistical physics.
\end{abstract}

\section{Introduction}

Learning is intrinsically a multilevel process. In both biological and artificial systems, this process is defined through a web of design choices that can steer the learning trajectory toward crucially different outcomes. In machine learning (ML), this multilevel structure underlies the optimization pipeline: model parameters are adjusted by a learning algorithm, e.g., stochastic gradient descent (SGD), that itself depends on a set of higher‐order decisions, specifying the network architecture, hyperparameters, and data‐selection procedures. These metaparameters are often adjusted dynamically throughout training following predefined schedules to enhance performance.  Biological learning is also mediated by a range of control signals across scales. Cognitive control mechanisms are known to modulate attention and regulate learning efforts to improve flexibility and multi-tasking \cite{botvinick14cognitivecontrol}. Additionally, structured training protocols are widely adopted in animal and human training to make learning processes faster and more robust \cite{liu2008easy}.

Optimizing the training schedules, effectively ``learning to learn,'' is a crucial problem in ML. However, the proposed solutions remain largely based on trial-and-error heuristics and often lack a principled assessment of their optimality. The increasing complexity of modern ML architectures has led to a proliferation of metaparameters, exacerbating this issue. As a result, several paradigms for automatic learning, such as meta-learning and hyperparameter optimization \cite{hutter2019automated}, have been developed. Proposed methods range from grid and random hyperparameter searches \cite{bergstra2012random} to Bayesian approaches \cite{snoek2012practical} and gradient‐based meta‐optimization \cite{maclaurin2015gradient,finn2017model}. However, these methods operate in high‐dimensional, nonconvex search spaces, making them computationally expensive and often yielding strategies that are hard to interpret. Although one can frame the selection of training protocols as an optimal‐control (OC) problem, applying standard control techniques to the full parameter space is often infeasible due to the curse of dimensionality.

Statistical physics provides theoretical methods to extract low-dimensional effective descriptions of high-dimensional learning problems in terms of order parameters that capture the key properties of training and performance \cite{engel2001statistical}. Several results have been obtained in the Bayes-optimal setting, characterizing the information-theoretically optimal performance for given data-generating processes and providing a threshold that no algorithm can improve \cite{doi:10.1073/pnas.1802705116}. In parallel, the algorithmic performance of practical procedures, such as empirical risk minimization, has been studied both in the asymptotic regime \cite{loureiro2021learning,pmlr-v119-mignacco20a,pmlr-v119-gerace20a} and through explicit analyses of training dynamics \cite{goldt2019dynamics,mignacco2020dynamical,bordelon2022self}. More recently, neural network models analyzed with statistical physics methods have been used to study various paradigmatic learning settings relevant to cognitive science \cite{NEURIPS2022_84bad835,leeanimals,strittmatter_mannelli_ruiz-garcia_musslick_spitzer_2025}. However, these lines of work have mainly focused on predefined protocols, often keeping metaparameters constant during training, without addressing the derivation of optimal learning schedules.

In this paper, we propose a unified framework for optimal learning that combines statistical physics and control theory to systematically identify training schedules across a broad range of learning scenarios. Specifically, we define an OC problem directly on the low-dimensional dynamics of the order parameters, where the metaparameters of the learning process serve as controls and the final performance is the objective. This approach serves as a testbed for uncovering general principles of optimal learning and offers two key advantages. First, the reduced descriptions of the learning dynamics circumvent the curse of dimensionality, enabling the application of standard control-theoretic techniques. Second, the order parameters capture essential aspects of the learning dynamics, allowing for interpretable outcomes. This formulation enables a unified treatment of diverse learning paradigms and their associated metaparameter schedules, such as task ordering, learning rate tuning, and dynamic modulation of the node activations. A variety of learning constraints and control budgets can be directly incorporated.

\section{Theoretical framework}
\label{sec:theory}

\subsection{The model}
\label{sec:model}
We study a general learning setting based on the \emph{sequence multi-index model} introduced in \cite{cui2024high}. This model captures a broad class of learning scenarios, both supervised and unsupervised, and admits a closed-form analytical description of its training dynamics.

We consider a dataset
$\mathcal{D} = \bigl\{(\bm{x}^\mu,y^\mu)\bigr\}_{\mu=1}^P$ of $P$ samples,
where $\bm{x}^\mu\in\mathbb{R}^{N\times L}$ are i.i.d.\ inputs and $y^\mu\in\mathbb{R}$ are the corresponding labels (if supervised learning is considered). Each input sample ${\bm x}\in \mathbb{R}^{N\times L}$, a sequence with $L$ elements ${\bm x}_{l}$ of dimension $N$, is drawn from a Gaussian mixture
\begin{equation}
    {\bm x}_{l}\sim \mathcal{N}\left(\frac{{\bm \mu}_{l,c_{l}}}{\sqrt N},\sigma^2_{l,c_{l}}\bm{I}_N\right)\,,\label{eq:distribution_input}
\end{equation}
where $c_{l}\in \{1\,,\ldots\,,C_{l}\}$ denotes cluster membership\footnote{Note that $\mu$ is used both as a sample index and to denote the mean vectors ${\bm \mu}_{l,c_l}$ of the Gaussian mixture. The distinction will be clear from the subscript.}. We emphasize that while we specify a Gaussian mixture here, our theoretical results remain valid for arbitrary distributions due to the Central Limit Theorem, provided that the variance $\sigma^2_{l,c_{l}}$ is finite. The random vector ${\bm c}=\{c_{l}\}_{l=1}^L$ is sampled from a probability distribution $p_c({\bm c})$, which can encode arbitrary correlations. In supervised settings, we will often assume 
\begin{equation}
    y=f^*_{{\bm w}_*}({\bm x})+\sigma_n z,\qquad
z\sim\mathcal{N}(0,1), \label{eq:label_def}
\end{equation}
where $f^*_{{\bm w}_*}({\bm x})$ is a fixed \emph{teacher} network with $M$ hidden units and parameters ${\bm w}_*\in \mathbb{R}^{N\times M}$, and \(\sigma_{n}\) controls label noise. This setting is known as \emph{teacher–student} (TS) \cite{engel2001statistical}.

We consider a two-layer neural network $f_{\bm w,\bm v}(\bm x)=\tilde f\bigl(\tfrac{\bm x^{\top}\,\bm w}{\sqrt{N}},\bm v\bigr)$ with $K$ hidden units. In a TS setting, this network serves as the \emph{student}. The parameters $\bm w\in\mathbb{R}^{N\times K}$ (first-layer) and $\bm v\in\mathbb{R}^{K\times H}$ (readout) are both trainable. The readout $\bm v$ has $H$ heads, $\bm{v}_h\in\mathbb{R}^K$ for $h=1,\dots,H$, which can be switched to adapt to different contexts or tasks. In the simplest case, $H=L=1$, the network will often take the form 
\begin{equation}
    f_{\bm w,\bm v}(\bm{x})=\frac{1}{\sqrt{K}}\sum_{k=1}^K v_{k} ~g\left(\frac{{\bm w}_k \cdot {
    \bm x}    }{\sqrt{N}} \right) \,,
\end{equation}
where we have dropped the head index, and $g(\cdot)$ is a nonlinearity (e.g., $g(z)=\operatorname{erf}(z/\sqrt{2}))$.

To characterize the learning process, we consider a cost function of the form
\begin{equation}
\mathcal{L}=\ell\left(\frac{{\bm x}^{\top} {\bm w_*}}{\sqrt{N}},\frac{{\bm x}^{\top}{\bm w}}{\sqrt{N}},\frac{\ww^\top\ww}{N} ,{\bm v}, {\bm c},z\right) + \tilde{g}\left(\frac{\ww^\top\ww}{N},{\bm v}\right)\,,\label{def:empirical_risk}
\end{equation}
where we have introduced the loss function $\ell$, and the regularization function $\tilde g$. Note that the functional form of $\ell(\cdot)$ in \eqref{def:empirical_risk} implicitly contains details of the problem, including the network architecture, the specific loss function used, and the target function. Additionally, it may contain adaptive hyperparameters and controls on architectural features. When considering a TS setting, the loss takes the form
\begin{equation}
    \ell\left(\frac{{\bm x}^{\top} {\bm w_*}}{\sqrt{N}},\frac{{\bm x}^{\top}{\bm w}}{\sqrt{N}},\frac{\ww^\top\ww}{N} ,{\bm v}, {\bm c},z\right) =\tilde{\ell}(f_{\bm w,\bm v}(\bm x),y)\,,\label{eq:loss_specialTS}
\end{equation}
where $y$ is given in \eqref{eq:label_def}. A typical choice is the square loss: $\tilde{\ell}(a,b)=(a-b)^2/2$.

\subsection{Learning dynamics}
\label{sec:learning_dynamics}

We study the learning dynamics under online SGD, in which each update is computed using a fresh sample $\bm{x}^\mu$ at each training step $\mu$. The parameters evolve as
\begin{align}
{\bm w}^{\mu+1}={\bm w}^{\mu}-{\eta}\nabla_{\ww} \mathcal{L}\;,\quad    \vv^{\mu+1}=\vv^\mu -\frac{\eta_v}{N}\nabla_{\vv} \mathcal{L}\;,
\end{align}
where $\eta$ and $\eta_v$ denote the learning rates. We take the limit where the input dimension $N$ and the number of training steps $\mu$ jointly tend to infinity at fixed training time $\alpha=\mu/N$. Note that, while the theory is derived in the high-dimensional limit $N \to \infty$, the agreement with finite-$N$ simulations is already excellent for moderate values of $N$ (see the Supplementary Material). Moreover, many real-world datasets of practical interest are inherently high-dimensional, making this regime broadly relevant. All other dimensions ($K$, $H$, $L$, and $M$) are assumed to be $\mathcal{O}_N(1)$. The generalization error is given by
\begin{equation}
    \epsilon_g ({\bm w},{\bm v}) = \mathbb{E}_{\bm x,\bm c}\left[ \ell_g\left(\frac{{\bm x}^{\top} {\bm w_*}}{\sqrt{N}},\frac{{\bm x}^{\top}{\bm w}}{\sqrt{N}},\frac{\ww^\top\ww}{N} ,{\bm v}, {\bm c},0\right) \right]\,,\label{eq:def_main_generror}
\end{equation}
where $\mathbb{E}_{\bm x,\bm c}$ denotes the expectation over the joint distribution of $\bm{x}$ and $\bm{c}$. Depending on the context, the function $\ell_g$ may coincide with the training loss $\ell$, or it may represent a different metric, such as the misclassification error in the case of binary labels. Crucially, the generalization error $\epsilon_g ({\bm w},{\bm v})$ depends on the first-layer weights only through the following low-dimensional \emph{order parameters}:
\begin{align}
\label{eq:main_def_orderpar}
\begin{split}
&Q^\mu_{kk'}\coloneqq \frac{{\ww^\mu_k}\cdot \ww^\mu_{k'}}{N}\;, \quad M^\mu_{km}\coloneqq\frac{{\ww^\mu_k}\cdot \ww_{*,m}}{N}\;,\quad\\ &R^\mu_{k(l,c_l)}\coloneqq\frac{{\ww^\mu_k}\cdot\mmu_{l,c_l}}{{N}}\;.
\end{split}
\end{align}
Collecting these together with the readout parameters $\bm v^\mu$ into a single vector $\mathbb{Q}=\left(\vec{\QQ}, \vec{\MM},\right.$ $\left.\vec{\RR}, \vec{\vv}\right)^\top$, we can write $\epsilon_g ({\bm w},{\bm v}) =\epsilon_g (\mathbb{Q})$ (see the Supplementary Material). Additionally, it is useful to define the low-dimensional constant parameters
\begin{align}
\begin{split}
& S_{m(l,c_l)}\coloneqq\frac{{\ww_{*,m}}\cdot\mmu_{l,c_l}}{{N}}\;,\quad
T_{mm'}\coloneqq \frac{{\ww_{*,m}}\cdot \ww_{*,m'}}{N}\;,\quad\\
& \Omega_{(l,c_l)(l',c'_{l'})}=\frac{\mmu_{l,c_l}\cdot\mmu_{l',c'_{l'}}}N\;. \label{eq:additional_params}
\end{split}
\end{align}
The scaling of teacher vectors $\ww_{*,m}$ and the centroids $\mmu_{l,c_l}$ is chosen so that the parameters in \eqref{eq:additional_params} are $\mathcal{O}_N(1)$.

In the high‐dimensional limit, the stochastic fluctuations of the order parameters $\mathbb{Q}$ vanish and their dynamics concentrate on a deterministic trajectory. Consequently, $\mathbb{Q}(\alpha)$ satisfies a closed system of ordinary differential equations (ODEs) \cite{biehl1995learning,saad1995PRL,goldt2019dynamics}:
\begin{align}
    \frac{\dd\mathbb{Q}}{\dd\alpha}=f_{\mathbb{Q}}\left(\mathbb{Q}(\alpha),\uu(\alpha)\right)\;, &&\alpha\in (0,\alpha_F]\;,\label{eq:ODE_compact}
\end{align}
with initial condition $\mathbb Q(0)=\mathbb Q_0$. Here, $\alpha_F=P/N$ denotes the final training time. The explicit form of $f_{\mathbb{Q}}$ is provided in the Supplementary Material, where we show that the theoretical predictions are in excellent agreement with numerical simulations. The vector $\bm u(\alpha)\in\mathcal{U}$ encodes the control parameters involved in the training process. The set of feasible controls $\mathcal U$ may include discrete, continuous, or mixed controls. For example, setting $\bm u(\alpha)=\eta(\alpha)$ corresponds to learning-rate schedules. Several specific examples are discussed in Section~\ref{sec:special_cases}.

\subsection{Optimal control}
\label{sec:optimal_control}
We seek to identify the OC $\bm u $ that minimizes the generalization error at the end of training. Note that, while we consider globally optimal schedules, previous works have also explored locally optimal schedules, maximizing the error decrease at each training step. Locally optimal approaches can significantly improve over constant schedules, however, they do not generally guarantee optimality of the final performance \cite{rattray1998analysis}. Furthermore, although our focus is on minimizing the final error, this framework can accommodate alternative objectives. For instance, one may optimize the time‐averaged generalization error as in \cite{carrasco2023meta}, if the performance during training is of interest. We adopt two types of OC techniques: \emph{indirect methods}, which solve the boundary‐value problem defined by the Pontryagin maximum principle \cite{kirk2004optimal}, and \emph{direct methods}, which discretize the control $\bm u(\alpha)$ and map the problem into a finite‐dimensional nonlinear program \cite{betts2010practical}. Additional costs or constraints associated with the control signal ${\bm u}$ can be directly incorporated into both classes of methods. For details on the OC implementation, see the Supplementary Material.

\subsection{Special cases of interest}
\label{sec:special_cases}

The proposed approach can be readily applied to describe several representative learning scenarios. Below we list some examples.

\subsubsection*{Hyperparameter schedules}

\paragraph{Learning rate.}
The learning rate $\eta$ is often regarded as the single most important hyperparameter \cite{bengio2012practical}. A small $\eta$ mitigates data noise but slows convergence, whereas a large $\eta$ accelerates convergence but amplifies fluctuations, which can lead to divergence. Several heuristic schedules have been proposed \cite{kalra2024warmup,loshchilov2016sgdr}, as well as gradient methods to optimize $\eta$ \cite{baydin2017online}. Optimal learning rate schedules were investigated in the 1990s in two-layer networks, using a variational approach closely related to ours \cite{saad1997globally,rattray1998analysis,schlosser1999optimization}. More recently, optimal learning rate schedules were derived in nonconvex optimization problems \cite{d2022optimal} and online continual learning \cite{mori2024optimal}. We further discuss optimal learning rates in the context of curriculum learning in Section~\ref{sec:applications_cl}.

\paragraph{Batch size.} Dynamically adjusting the batch size, i.e., the number of data samples used to estimate the gradient at each SGD step, has been proposed as a powerful alternative to learning rate schedules \cite{smith2017don}. Optimal mini-batch schedules can be derived within our formulation, as done in Section~\ref{sec:DAE} for batch augmentation to train a denoising autoencoder.

\subsubsection*{Dynamic data selection}

\paragraph{Task ordering.} The ability to learn new tasks without forgetting previously learned ones is crucial for both artificial and biological learners. Recent theoretical studies have assessed the relative effectiveness of various pre‐specified task sequences \cite{lee2021continual,evron2022catastrophic,shan2024order} within minimal models. In contrast, our framework allows to identify optimal task‐ordering strategies that minimize forgetting. In particular, the model in \cite{lee2021continual} is a special case of our formulation where each of the teacher vectors defines a different task $y_m=f^*_{\bm{w}^*_m}(\bm{x})$, $m=1,\ldots,M$, and $L=1$, the student has $K=M$ hidden nodes and $H=M$ task-specific readout heads. The task index $m$ can then be treated as a control variable to identify optimal ordering protocols that minimize the error across tasks \cite{mori2024optimal}.

\paragraph{Curriculum learning.} For datasets involving heterogeneous sample difficulty, it is natural to ask whether training performance can be enhanced by a curriculum, i.e., by presenting examples in a structured order based on their difficulty, rather than random sampling. This question has been explored in recent literature
\cite{NEURIPS2022_84bad835,NEURIPS2023_4c8ce3c6} and is investigated within our formulation in Section~\ref{sec:applications_cl}.

\paragraph{Data imbalance.}
Many real-world datasets exhibit class imbalance, with certain classes significantly under-represented. Recent work has studied class-imbalance mitigation through under- and over-sampling in sequential data \cite{loffredo2024restoring2}. Further aspects of data imbalance, such as relative representation imbalance and different sub-population variances, have been explored in \cite{mannelli2024bias,NEURIPS2024_2ba538ff}. These types of imbalance can be incorporated in our formulation, e.g., by tilting the distribution of cluster memberships $p_c(\bm{c})$, the cluster variances, and the alignment parameters $\bm{S}$ between teacher vectors and cluster centroids. This approach allows to investigate balance-restoring strategies such as optimal data ordering, adaptive loss reweighting, and learning-rate schedules.

\subsubsection*{Dynamic architectures}

\paragraph{Dropout.}Dropout is a widely adopted regularization technique in which random subsets of the network are deactivated during training to encourage robust feature representations \cite{srivastava2014dropout}. While empirical studies have proposed adaptive dropout probabilities to enhance performance \cite{morerio2017curriculum,liu2023dropout}, a theoretical understanding of optimal dropout schedules remains limited. We recently introduced a two‐layer network model incorporating dropout and analyzed the impact of fixed dropout rates \cite{mori2025analytic}. As shown in Section~\ref{sec:dropout}, our formulation contains the model of \cite{mori2025analytic} as a special case, enabling the derivation of principled dropout schedules.

\paragraph{Gating.}Gating functions modify the network architecture by selectively activating specific pathways, thereby modulating information flow based on input context. This principle improves model efficiency and expressiveness, and underlies diverse systems such as mixture of experts \cite{shazeer2017outrageously}, gated recurrent units \cite{gru} and gated linear networks \cite{veness2021gated}. The latter are analytically tractable and have been investigated in several theoretical works \cite{li2022globally,saxe2022neural,mignacco2025nonlinear}.
Our framework offers the possibility to study dynamic gating and adaptive modulation by controlling the hyperparameters of the gating functions. In TS settings, the model in \cite{mignacco2025nonlinear} arises as a special case of our formulation, where $L=1$ and $f_{\bm{w},\bm{v}}(\bm{x})=\sum_{k=1}^{\lfloor K/2 \rfloor}(\bm{w}_{\lfloor K/2 \rfloor+k}\cdot \bm{x}) g_k(\bm{w}_{k}\cdot \bm{x})$ with gating functions $g_k$.

\paragraph{Dynamic attention.} Self-attention is the core building block of transformer architectures \cite{vaswani2017attention}. Dynamic attention mechanisms enhance standard attention by adapting its structure in response to input properties or task requirements, for example, by selecting sparse token interactions \cite{roy2021efficient}, varying attention spans \cite{sukhbaatar2019adaptive}, or pruning attention heads \cite{michel2019sixteen}. Recent work introduced minimal models of dot‐product attention that admit an analytic characterization \cite{cui2024high}. A multi-head single-layer attention model can be recovered by setting
\begin{align}
f_{\bm{w},\bm{v}}(\bm{x})=\sum_{h=1}^H v^{(h)} \bm{x} \,{\tilde\sigma}\left(\frac{\bm{x}^\top \bm{w}^{(h)}_{\mathcal Q} {\bm{w}^{(h)}_{\mathcal K}}^\top\bm{x}}{N}\right)\in\mathbb{R}^{N\times L}\;,\label{eq:attention_block}
\end{align}
where ${\tilde \sigma}=\operatorname{softmax}$, $\bm{w}^{(h)}_{\mathcal{Q}}\in\mathbb{R}^{N\times D_H}$ and $\bm{w}^{(h)}_{\mathcal{K}}\in\mathbb{R}^{N\times D_H}$ denote the query and key matrices for the $h^{\rm th}$ head, with head dimension $D_H$. In TS settings, \eqref{eq:attention_block} is a special case of our formulation (see also \cite{cui2024high}). Possible controls include masking variables that dynamically prune attention heads, sparsify token interactions, or modulate context visibility.

\section{Results}
\label{sec:applications}

In this section, we present three learning scenarios in which we identify optimal learning strategies.

\subsection{Curriculum learning}
\label{sec:applications_cl}

Curriculum learning (CL) refers to a variety of training protocols in which examples are presented in a curated order—typically organized by difficulty or complexity. In animal and human training, CL is widely used and extensively studied in behavioral research, demonstrating clear benefits \cite{elio1984effects,pashler2013does}. By contrast, results on the efficacy of CL in machine learning remain sparse and less conclusive \cite{bengio2009curriculum,wang2021survey}. Empirical studies across diverse settings have nonetheless demonstrated that curricula can outperform standard heuristic strategies \cite{hacohen2019power,wu2020when}. Several theoretical studies have explored the benefits of curriculum learning in analytically tractable models. Easy-to-hard curricula have been shown to accelerate learning in convex settings \cite{weinshall2020theory,NEURIPS2022_84bad835} and improve generalization in nonconvex problems, such as XOR classification \cite{mannelli2024tilting} or parity functions \cite{NEURIPS2023_4c8ce3c6}. However, these analyses focused on predefined heuristics, which may not be optimal. Moreover, although hyperparameter schedules have been shown to enhance curriculum learning empirically \cite{pmlr-v130-zhou21a}, a principled approach to their joint optimization remains largely unexplored.

We focus on a prototypical model of CL introduced in
\cite{bengio2009curriculum} and recently studied analytically in \cite{NEURIPS2022_84bad835}. This model considers a binary classification problem in a TS setting where both teacher and student are single-layer networks. The input vectors consist of $L=2$ elements: relevant directions $\bm x_1$, which a single-layer teacher uses to generate labels $y=\operatorname{sign}({\bm x}_1\cdot{\bm w}_{*}/\sqrt{N})$, and irrelevant directions $\bm x_2$, which do not affect the labels. The student network is given by
\begin{equation}
    f_{\bm w}(\bm x)=\operatorname{erf}\left(\frac{{\bm x}_1\cdot {\bm w}_{1}+{\bm x}_2\cdot {\bm w}_{2}}{2\sqrt{N}}\right)\,.
\end{equation}
As a result, the student does not know a priori which directions are relevant. All inputs are single-cluster zero-mean Gaussian variables and the sample difficulty is controlled by the variance $\Delta$ of the irrelevant directions, while the relevant directions are assumed to have unit variance. We consider the squared loss $\ell=(y-f_{\bm w}(\bm x))^2/2$ and ridge regularization $\tilde{g}\left(\bm{w}^\top\bm{w}/N\right)=\lambda \left({\ww}_1 \cdot {\ww}_1+{\ww}_2 \cdot {\ww}_2\right)/(4N)$, with tunable strength $\lambda\geq 0$. Full expressions for the ODEs governing the learning dynamics of the order parameters $M_{11}={\bm w}_*\cdot{\bm w}_1/N$, $Q_{11}={\bm w}_1\cdot{\bm w}_1/N$, $Q_{22}={\bm w}_2\cdot{\bm w}_2/N$, and the generalization error are provided in the Supplementary Material.

\begin{figure}[t!]
    \centering
    \includegraphics[width=\linewidth]{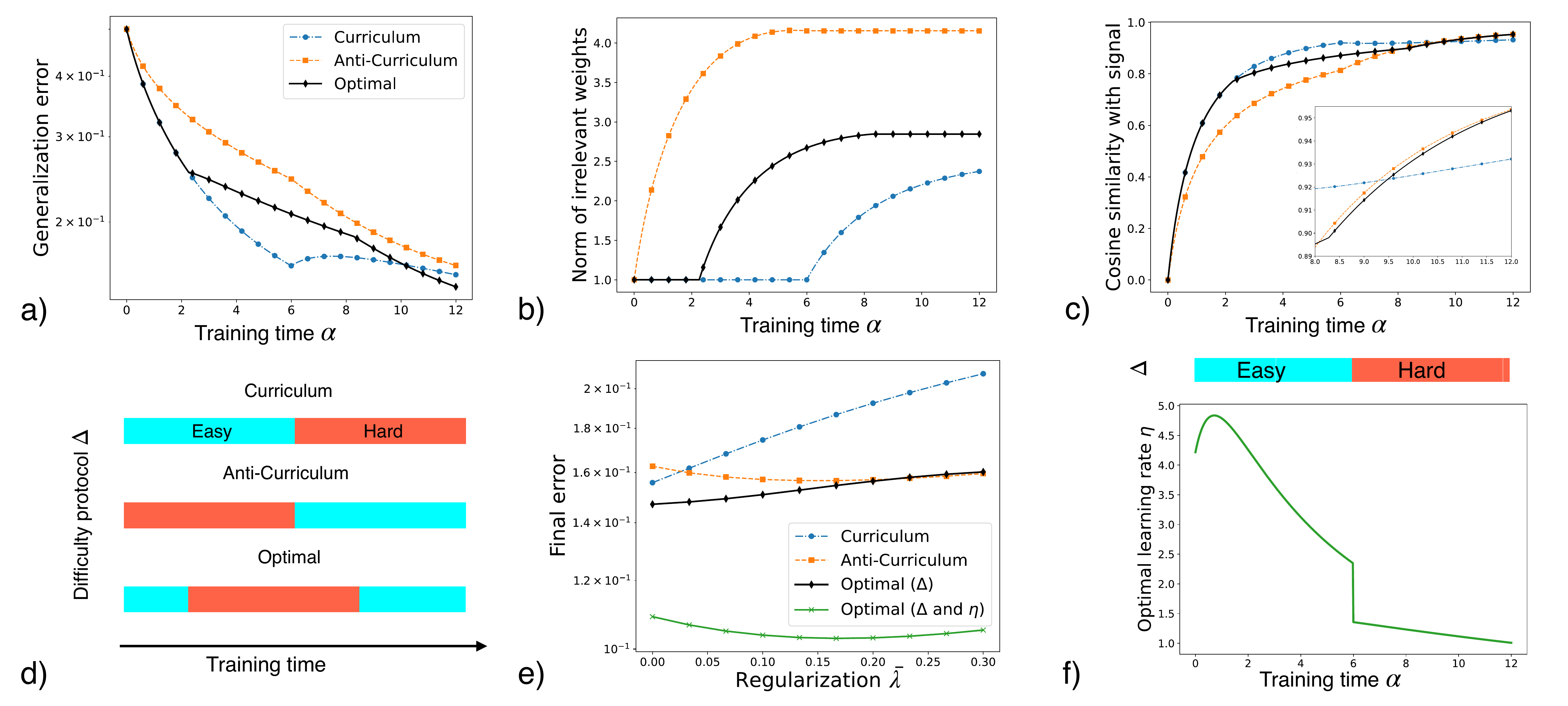}
    \caption{\textbf{Optimal curricula compared to benchmarks.}
    \textbf{a)} Generalization error vs.~training time $\alpha$ for curriculum (easy-to-hard), anti-curriculum (hard-to-easy) and the optimal schedule at constant learning rate.
    \textbf{b)} Squared norm of irrelevant weights $Q_{22}$.
    \textbf{c)} Cosine similarity with the target signal $M_{11}/\sqrt{T_{11} Q_{11}}$ (inset zooms into the late-training regime).
    \textbf{d)} Timeline of each schedule.
    {\bf e)} Generalization error at the final time $\alpha_F=12$ as a function of the (rescaled) regularization $\bar{\lambda}=\lambda\eta$ for the three strategies presented in panels a-c, obtained optimizing over $\Delta$ at constant $\eta=3$, and the optimal strategy obtained by jointly optimizing $\Delta$ and $\eta$ (displayed in panel {\bf f} for $\lambda=0$).
    {\bf Parameters: }$\alpha_F=12$, $\Delta_1=0$, $\Delta_2=2$, $T_{11}=2$. Panels a-d: $\eta=3$, $\lambda=0$. {\bf Initialization:} $Q_{11}=Q_{22}=1$, $M_{11}=0$.}
    \label{fig:curr_comparison}
\end{figure}

We consider a dataset composed of two difficulty levels: $50\%$ ``easy'' examples ($\Delta=\Delta_1$), and $50\%$ ``hard'' examples ($\Delta=\Delta_2>\Delta_1$). We call \emph{curriculum} the easy-to-hard schedule in which all easy samples are presented first, and \emph{anti-curriculum} the opposite strategy (see Figure~\ref{fig:curr_comparison}d). We compute the optimal sampling  strategy $\bm{u}(\alpha)=\Delta(\alpha)\in\{\Delta_1,\Delta_2\}$ using Pontryagin's maximum principle to minimize the final misclassification error averaged over an equal proportion of easy and hard examples.

Good generalization requires balancing two competing objectives: maximizing the teacher–student alignment along relevant directions, as measured by the cosine similarity with the signal $M_{11}/\sqrt{T_{11}Q_{11}}$, and minimizing the norm of the student’s weights along the irrelevant directions,
$\sqrt{Q_{22}}$. We observe that anti-curriculum favors the first objective, while curriculum the latter (see Figure~\ref{fig:curr_comparison}). In this case, the optimal strategy is non-monotonic in difficulty, following an ``easy-hard-easy'' schedule, that balances the two objectives (see panels \ref{fig:curr_comparison}b and \ref{fig:curr_comparison}c), and achieves lower generalization error compared to the two monotonic strategies.

The optimal balance between these competing goals is determined by the interplay between the difficulty schedule and other problem hyperparameters such as regularization and learning rate. Figure~\ref{fig:curr_comparison}e shows the final generalization error as a function of the regularization strength for curriculum (blue), anti-curriculum (orange), and the optimal schedule (black). At high regularization, weight decay alone ensures norm suppression along the irrelevant directions, so the optimal strategy reduces to anti-curriculum.

We next explore how a time-dependent learning rate $\eta(\alpha)$ can be coupled with the curriculum, by extending the control vector to $\uu=(\Delta,\eta)$, so that difficulty and learning-rate schedules are optimized jointly. In Figure~\ref{fig:curr_comparison}e, we see that this joint optimization produces a substantial reduction in generalization error compared to any constant‐$\eta$ strategy. Interestingly, for all parameter settings considered, an easy‐to‐hard curriculum becomes optimal once the learning rate is properly adjusted. Figure~\ref{fig:curr_comparison}f displays the optimal $\eta(\alpha)$ at $\lambda=0$: it begins with a warm‐up phase, transitions to gradual annealing, and then undergoes a sharp drop precisely when the curriculum shifts from easy to hard samples. This behavior is intuitive, since learning harder examples benefits from a lower learning rate. These results align with the empirical learning rate scheduling employed in the numerical experiments of \cite{mannelli2024tilting}. Importantly, our framework provides a principled derivation of the optimal joint schedule, thereby confirming and grounding prior empirical insights.

To validate these theoretical predictions beyond the analytically tractable setting, we test the easy-hard-easy schedule on a Cluttered CIFAR-10 classification task using a convolutional neural network (see the Supplementary Material for details). Consistent with the optimal control solution, we find that the non-monotonic easy-hard-easy strategy outperforms both the standard curriculum and anti-curriculum baselines, confirming that the qualitative structure of the optimal schedule transfers to realistic architectures and datasets.

\subsection{Dropout regularization}
\label{sec:dropout}

Dropout \cite{srivastava2014dropout} is a regularization technique designed to prevent harmful co-adaptations of hidden units, thereby reducing overfitting and enhancing the network's performance. During training, each node is independently kept active with probability $p$ and “dropped” (i.e., its output set to zero) otherwise, effectively sampling a random subnetwork at each iteration. At test time, the full network is used, which corresponds to averaging over the ensemble of all subnetworks and yields more robust predictions.  While early works recommended keeping the activation probability constant throughout training \cite{srivastava2014dropout}, recent empirical studies propose adaptive schedules to further enhance performance \cite{7078567,morerio2017curriculum}. In particular, \cite{morerio2017curriculum} showed that heuristic schedules that decrease the activation probability over time are akin to easy-to-hard curricula and can improve performance. Although adaptive dropout schedules have attracted practical interest, the conditions under which they outperform constant strategies remain poorly understood.

In \cite{mori2025analytic}, we introduced a prototypical model of dropout and derived analytic results for constant dropout probabilities. We showed that dropout reduces harmful node correlations---quantified via order parameters---and consequently improves generalization. We further demonstrated that the optimal (constant) activation probability decreases as the variance of the label noise increases. In this section, we extend this analysis to optimal dropout schedules.

\begin{figure}[t!]
    \centering
    \includegraphics[width=\linewidth]{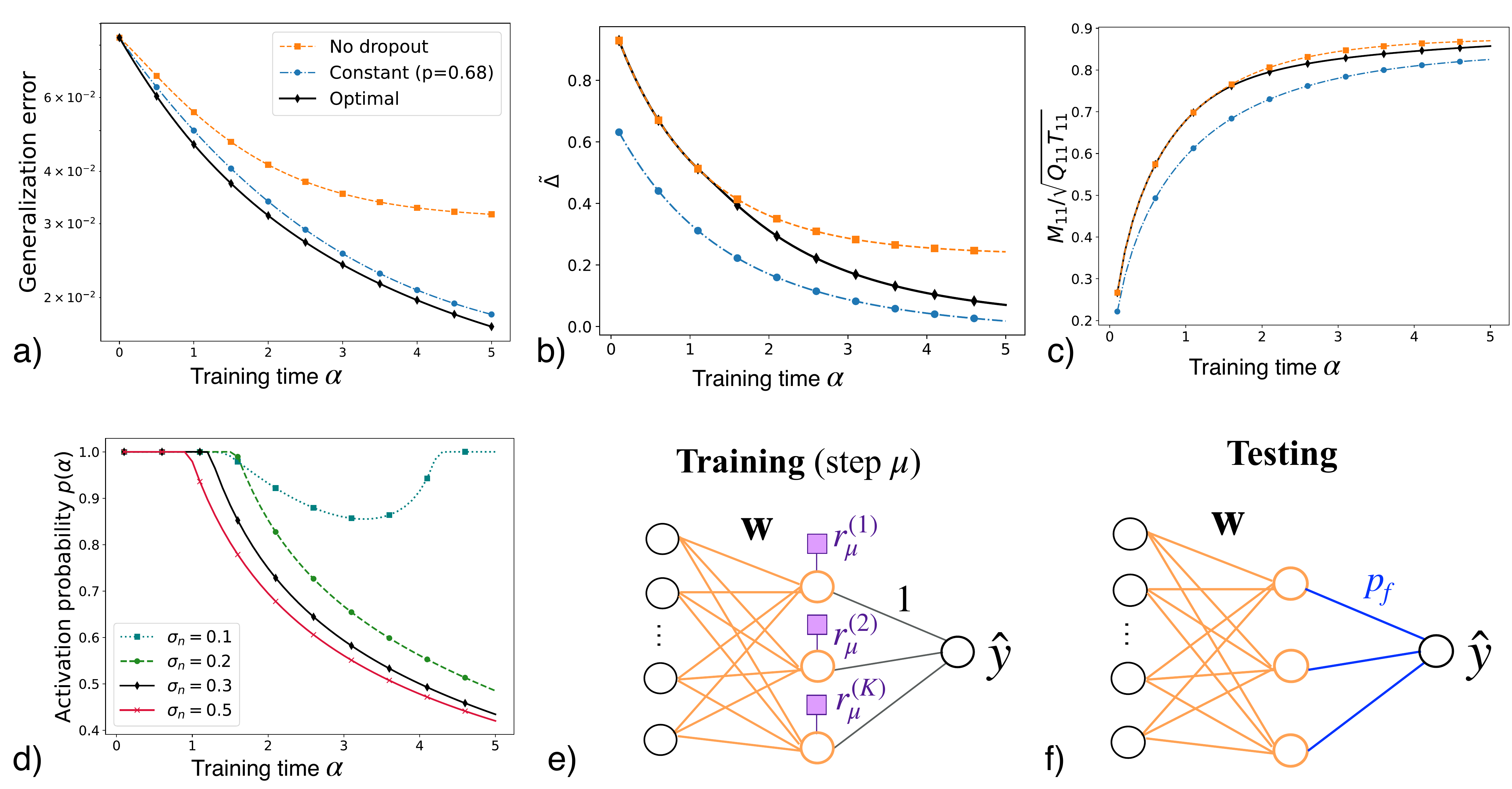}
    \caption{\textbf{Optimal dropout schedules.} \textbf{a)} Generalization error vs.\ training time $\alpha$ without dropout, for constant activation probability $p=p_f=0.68$, and for the optimal dropout schedule with $p_f=0.678$, at label noise $\sigma_n=0.3$. \textbf{b)} Detrimental correlations between the student's hidden nodes, measured by $\tilde\Delta=(Q_{12} - M_{11} M_{21}) / \sqrt{Q_{11} Q_{22}}$ at $\sigma_n=0.3$. \textbf{c)} Teacher-student cosine similarity $M_{11}/\sqrt{Q_{11}T_{11}}$ at $\sigma_n=0.3$. {\bf d)} Optimal dropout schedules for different label-noise levels. {\bf e}-{\bf f)} Dropout model at training and testing times. \textbf{Parameters:} $\alpha_F=5$, $K=2$, $M=1$, $\eta=1$. The teacher weights $\bm{w}^*$ are drawn i.i.d.~from $\mathcal{N}(0,1)$ with $N=10000$. The student weights are initialized to zero.}
    \label{fig:drop_comparison}
\end{figure}

We consider a TS setup where both teacher and student networks are soft-committee machines \cite{saad1995PRL}, i.e., two-layer networks with untrained readout weights set to one. The inputs $\bm x\in \mathbb{R}^N$ are taken to be standard Gaussian variables and the corresponding labels are 
\begin{align}
y=f^*_{\bm w_*}(\bm{x})+\sigma_n \,z\;,&&
    f^*_{\bm w_*}(\bm{x}) = \sum_{m=1}^M \operatorname{erf} \left(\frac{\bm w_{*,m}\cdot{\bm x}}{\sqrt{N}}\right)\,,
\end{align}
where $z\sim\mathcal{N}(0,1)$. To describe dropout, at each training step $\mu$ we couple i.i.d.~node-activation Bernoulli random variables $r^{(k)}_\mu\sim{\rm Ber}(p_\mu)$ to each of the hidden nodes (see Fig.~\ref{fig:drop_comparison}e):
\begin{equation}
    f^{\rm train}_{\bm w}(\bm{x}^\mu)=\sum_{k=1}^K r^{(k)}_\mu  \operatorname{erf} \left(\frac{\bm w_{k}\cdot{\bm x}^\mu}{\sqrt{N}}\right)\,,
\end{equation}
so that node $k$ is active if $r^{(k)}_\mu=1$. 
At testing time, the full network is used as (see Fig.~\ref{fig:drop_comparison}f)
\begin{equation}
    f^{\rm test}_{\bm w}(\bm{x})=\sum_{k=1}^K p_f  \operatorname{erf} \left(\frac{\bm w_{k}\cdot{\bm x}}{\sqrt{N}}\right)\,.
\end{equation}
The rescaling factor $p_f$ ensures that the reduced activity during training is taken into account when testing. We consider the squared loss $\ell=(y-f_{\bm w}(\bm x))^2/2$ and no weight-decay regularization. The ODEs governing the order parameters $M_{km}=\ww_k\cdot \ww_{*,m}/N$ and $Q_{jk}=\ww_k \cdot\ww_j/N$, as well as the resulting generalization error, are given in the Supplementary Material. 

For simplicity, we focus on the case $M=1$ and $K=2$, although our considerations hold more generally. During training, assuming $T_{11}=1$, each student weight vector can be decomposed as
${\bm w}_i=M_{i1}{\bm w}_{*,1}+\tilde{{\bm w}}_i$,
where $\tilde{\bm w}_i\perp\bm w_{*,1}$ denotes the uninformative component acquired due to input and label noise. Generalization requires balancing two competing goals: improving the alignment of each hidden unit with the teacher, measured by $M_{i1}$, and reducing correlations between their uninformative components, $\tilde{\bm w}_1$ and $\tilde{\bm w}_2$, so that noise effects cancel rather than compound. We quantify these detrimental correlations by the observable $\tilde\Delta=(Q_{12} - M_{11} M_{21}) / \sqrt{Q_{11} Q_{22}}$. Figure \ref{fig:drop_comparison}b compares a constant dropout strategy ($p=p_f=0.68$, orange) with no dropout ($p=p_f=1$, blue) and shows that dropout sharply reduces $\tilde\Delta$ during training. Intuitively, without dropout, both nodes share identical noise realizations at each step, reinforcing their uninformative correlation; with dropout, nodes are from time to time trained individually, reducing correlations. Although dropout also slows the growth of the teacher–student cosine similarity (Figure~\ref{fig:drop_comparison}c) by reducing the number of updates per node, the large decrease in $\tilde\Delta$ leads to an overall lower generalization error (Figure~\ref{fig:drop_comparison}a).

To find the optimal dropout schedule, we treat the activation probability as the control variable, $u(\alpha)=p(\alpha)\in[0,1]$. Additionally, we optimize over the final rescaling $p_f\in [0,1]$ to minimize the final error. We solve this OC problem using a direct multiple‐shooting method implemented in CasADi (see the Supplementary Material). Figure~\ref{fig:drop_comparison} shows the resulting optimal schedules for increasing label noise $\sigma_n$. Each schedule exhibits an initial period with no dropout ($p=1$) followed by a gradual decrease of $p(\alpha)$. These strategies resemble those heuristically proposed in \cite{morerio2017curriculum} but are obtained here via a principled procedure. 

The order parameters of the theory suggest a simple interpretation of the optimal schedules. In the initial phase of training, it is beneficial to fully exploit the rapid increase in the teacher-student cosine similarity by keeping both nodes active (see Figure~\ref{fig:drop_comparison}). Once  the increase in cosine similarity plateaus, it becomes more advantageous to decrease the activation probability in order to mitigate negative correlations among the student’s nodes.

Noisier tasks induce stronger detrimental correlations and therefore require a lower activation probability. Indeed, Figure~\ref{fig:drop_comparison}d shows that, as $\sigma_n$ grows, the initial no‐dropout phase becomes shorter and the activation probability decreases more sharply. Conversely, at low label noise, the activation probability remains close to one.

\subsection{Denoising autoencoder}
\label{sec:DAE}

\begin{figure}[t!]
    \centering
    \includegraphics[width=\linewidth]{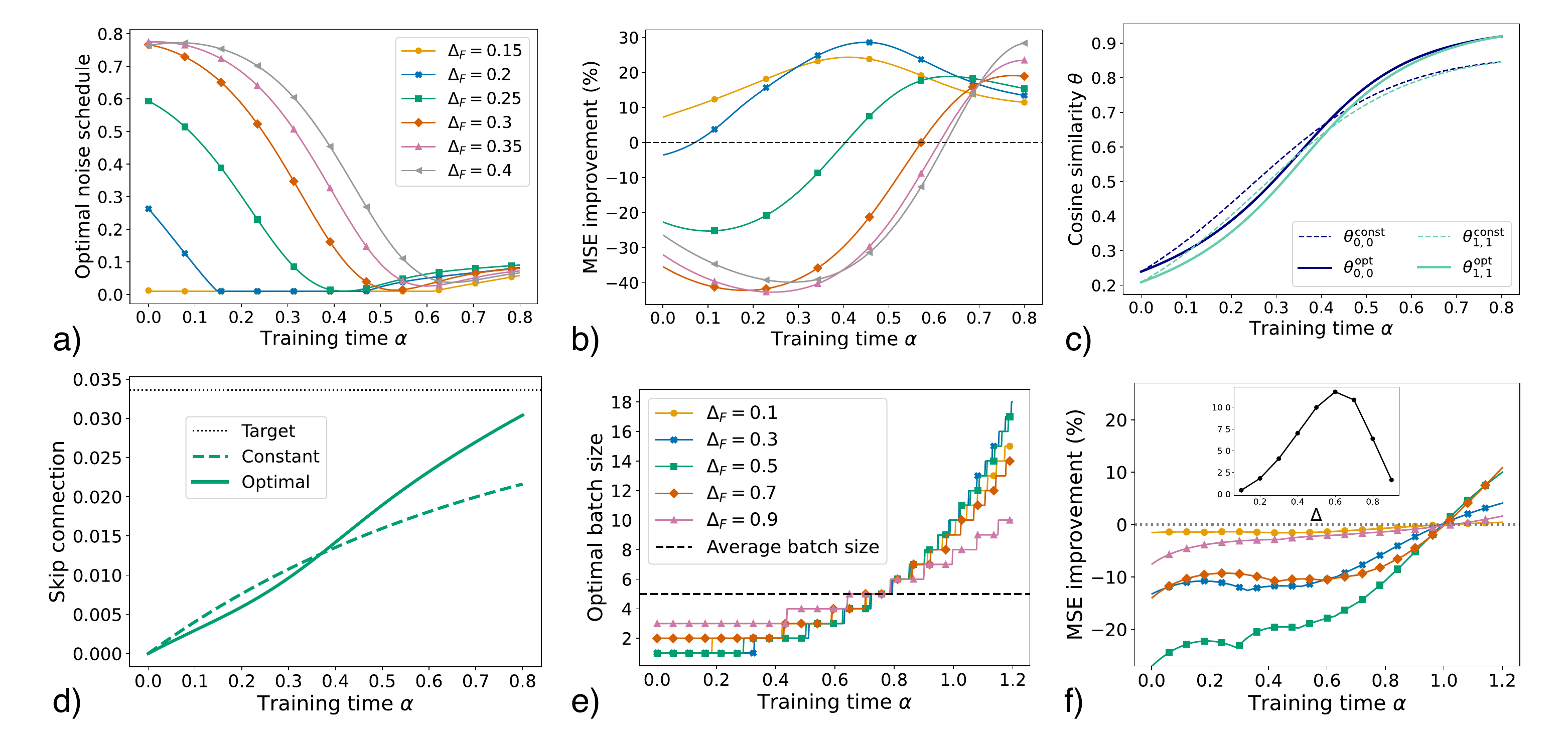}
    \caption{\textbf{Optimal schedules for denoising autoencoders.} \textbf{a)} Optimal noise schedule $\Delta$ vs.\ training time $\alpha$ for different values of the test noise level $\Delta_F$. \textbf{b)} Percentage MSE improvement of the optimal noise strategy compared to the constant one at $\Delta(\alpha)=\Delta_F$, computed as: $100 (\operatorname{MSE}_{\rm const}(\alpha)-\operatorname{MSE}_{\rm opt}(\alpha))/(\operatorname{MSE}_{\rm const}(0)-\operatorname{MSE}_{\rm const}(\alpha))$. \textbf{c)} Cosine similarity $\theta_{k,k}=R_{k(1,k)}/\sqrt{Q_{kk}\Omega_{(1,k)(1,k)}}$ for the optimal noise schedule (full lines) and the constant schedule (dashed lines), at $\Delta_F=0.25$. \textbf{d)} Skip connection $b$ for the optimal noise schedule and the constant one at $\Delta_F=0.25$. The dotted line marks the target value $b^{*}$ given by \eqref{eq:optimal_skipconnection}. \textbf{e)} Optimal batch augmentation schedule for different values of $\Delta_F$. All schedules have average batch size $\bar{B}=5$. \textbf{f)} MSE improvement of the optimal batch strategy compared to the constant one at $B(\alpha)=\bar{B}=5$. Inset: MSE improvement at $\alpha_F=1.2$ vs.~$\Delta$. {\bf Parameters:} $K=C_1=2$, $\eta=5$, $\sigma=0.1$, $N=1000$, $g(z)=z$. The skip connection $b$ is trained ($\eta_b=5$, $b(0)=0$) in panels {a-d} and fixed to $b^*$ in {e-f}. Initial conditions are given in the Supplementary Material.}
    \label{fig:DAEschedule}
\end{figure}

Denoising autoencoders (DAEs) are neural networks trained to reconstruct input data from their corrupted version, thereby learning robust feature representations \cite{10.1145/1390156.1390294}. Recent developments in diffusion models have revived the interest in denoising tasks as a key component of the generative process \cite{NEURIPS2020_4c5bcfec}. Several theoretical works have studied the learning dynamics and generalization properties of DAEs \cite{pretorius2018learning,cui2023high,cui2025precise}.

A series of empirical works have considered noise schedules in the training of DAEs. \cite{GerasSutton2015ScheDA} showed that adaptive noise levels promote learning multi-scale representations. Similarly, in diffusion models, networks are trained to denoise inputs at successive diffusion timesteps, each linked to a specific noise level. Recent work \cite{zheng2024non} demonstrates that non-uniform sampling of diffusion time, effectively implementing a noise schedule, can further enhance performance. Additionally, data augmentation, where multiple independent corrupted samples are obtained for each clean input, is often employed \cite{chen2014marginalized}. However, identifying principled noise schedules and data augmentation strategies remains largely an open problem. In this section, we address these questions within the model introduced in \cite{cui2023high}.

We consider input data $\bm{x}=(\bm{x}_1,\bm{x}_2)\in\mathbb{R}^{N\times 2}$, where $\xx_1\sim\mathcal{N}\left(\frac{\mmu_{1,c_1}}{\sqrt N},\sigma_{1,c_1}^2\bm{I}_N\right)$, $c_1=1,\ldots,C_1$, represents the clean input drawn from a Gaussian mixture of $C_1$ clusters, while $\bm{x}_2\sim\mathcal{N}(\bm{0},\bm{I}_N)$ is additive standard Gaussian noise. We will take $\sigma_{1,c_1}=\sigma$ for all $c_1$ and equiprobable clusters unless otherwise stated. The network receives the noisy input $\tilde{\bm{x}}=\sqrt{1-\Delta}\,\bm{x}_1 +\sqrt{\Delta}\,\bm{x}_2$, where $\Delta>0$ controls the level of corruption. The denoising is performed via a two-layer autoencoder
\begin{align}
    f_{\bm{w},b}(\tilde{\bm{x}})= \frac{\bm{w}}{\sqrt{N}}g\left(\frac{\bm{w}^\top\tilde{\bm{x}}}{\sqrt{N}}\right)+ b \,\tilde{\bm{x}}\;\in\mathbb{R}^N\;,
\end{align}
with tied weights $\bm{w}\in\mathbb{R}^{N\times K}$, with hidden dimension $K$ and a skip connection $b\in\mathbb{R}$. The activation function $g$ is applied component-wise. The loss function is given by the squared reconstruction error: $ \mathcal{L}(\ww,b|\bm{x},\bm{c})=\Vert \xx_1-f_{\ww,b}(\tilde{\xx})\Vert_2^2/2$. This loss can be recast in the form of \eqref{def:empirical_risk}, as shown in \cite{cui2024high}. The skip connection is trained via online SGD, i.e., $b^{\mu+1}=b^\mu -(\eta_b/N)\partial_b\mathcal{L}({\bm w}^\mu,b^\mu|{\bm x}^{\mu},{\bm c}^{\mu})$.

We measure generalization via the mean squared error: $\operatorname{MSE}=\mathbb{E}_{\bm{x},\bm{c}}\left[\Vert \xx-f_{\ww,b}(\tilde{\xx})\Vert_2^2/2\right]$. In the high-dimensional limit, the MSE is given by (see the Supplementary Material)
\begin{align}
\begin{split}
    &  \text{MSE}=N\left[\sigma^2\left(1-b\sqrt{1-\Delta}\right)^2+b^2\Delta\right]\\ & +\mathbb{E}_{\bm{x},\bm{c}}\left[\sum_{k,k'=1}^K Q_{kk'}  g(\tilde\lambda_k) g(\tilde\lambda_{k'})-2\sum_{k=1}^K (\lambda_{1,k}-b\tilde\lambda_k)g(\tilde\lambda_k)\right],\label{eq:MSE_dae}
    \end{split}
\end{align}
where we have defined the pre-activations $\tilde{\lambda}_k\equiv { \tilde{\bm x}}\cdot{\bm w}_k/\sqrt{N}$ and $\lambda_{1,k}= {\bm w}_k\cdot {\bm x}_1/\sqrt{N}$, and dropped a constant term. Note that the leading term in \eqref{eq:MSE_dae}---proportional to $N$---is independent of the autoencoder weights $\bm{w}$, and depends only on the skip connection $b$ and the noise level $\Delta$. Therefore, the presence of the skip connection can improve the MSE by a contribution of order $\mathcal{O}_N(N)$. To leading order, the optimal skip connection is 
 \begin{equation}
\label{eq:optimal_skipconnection}
    b^{*}=\frac{\sqrt{(1 - \Delta )} \, \sigma^2}{(1 - \Delta )\, \sigma^2 + \Delta } \;.
\end{equation}
 In the Supplementary Material, we provide closed-form expressions for the MSE and the ODEs describing the evolution of the order parameters.

We start by considering the problem of finding the optimal noise schedule $\Delta(\alpha)$ that minimizes the final MSE, computed at the fixed test noise level $\Delta_F$. We use a direct multiple-shooting method implemented in CasADi (see the Supplementary Material). In the following analysis, we consider linear activation. Figure~\ref{fig:DAEschedule}a displays the optimal noise schedules for a range of test noise levels $\Delta_F$. We observe that the optimal schedule typically features an initial decrease, followed by a moderate increase toward the end. At low $\Delta_F$, the optimal schedule remains nearly flat and close to $\Delta=0$ before the final increase. Both the duration of the initial decreasing phase and the average noise level throughout the schedule increase with $\Delta_F$. Figure~\ref{fig:DAEschedule}b shows that the optimal schedule improves the MSE by approximately $10$-$30\%$ over the constant schedule $\Delta(\alpha)=\Delta_F$. The optimal schedule achieves two key objectives. First, it enhances the reconstruction capability of the autoencoder network, leading to a higher cosine similarity between the hidden nodes and the means of the Gaussian mixture defining the clean input distribution (panel \ref{fig:DAEschedule}c). Second, it accelerates the convergence of the skip connection toward the target value $b^*$ in \eqref{eq:optimal_skipconnection} (panel \ref{fig:DAEschedule}d).

We then explore a setting that incorporates data augmentation, with inputs $\bm{x}=(\bm{x}_1, \bm{x}_{2}, \ldots, \bm{x}_{B+1})\in\mathbb{R}^{N\times B+1}$, where $\xx_1\sim\mathcal{N}\left(\frac{\mmu_{1,c_1}}{\sqrt N},\sigma^2\bm{I}_N\right)$ denotes the clean version of the input as before. We consider $B$ independent realizations of standard Gaussian noise $\bm{x}_{2}, \ldots, \bm{x}_{B+1}\overset{\rm i.i.d.}{\sim}\mathcal{N}(\bm{0},\bm{I}_N)$ and construct a batch of noisy inputs: $\tilde{\bm{x}}_a=\sqrt{1-\Delta}\,\bm{x}_1 +\sqrt{\Delta}\,\bm{x}_{a+1}$, $a=1,\ldots,B$. The loss is averaged over the batch: $ \mathcal{L}(\ww,b|\bm{x},\bm{c})=\sum_{a=1}^B\Vert \xx_1-f_{\ww,b}(\tilde{\xx}_a)\Vert_2^2/(2B)$. For simplicity, we take constant noise $\Delta=\Delta_F$ and we fix the skip connection to its optimal value $b^*$.

We study the optimal batch size schedule $B(\alpha)\in\mathbb{N}$ at fixed sample budget $B_{\rm tot}=\bar B \alpha_F N$, where $\bar B$ is the average batch size available at each training time. As shown in Figure~\ref{fig:DAEschedule}e, the optimal schedule features a progressive increase in batch size throughout training, with only a moderate dependence on $\Delta_F$. This corresponds to averaging the loss over a growing number of noise realizations, effectively reducing gradient variance and acting as a form of annealing that stabilizes learning in the later phases. This strategy leads to an MSE improvement of up to $10\%$ compared to the constant schedule preserving the total sample budget ($B(\alpha)=\bar B$), as depicted in Figure~\ref{fig:DAEschedule}f. The inset shows that the final MSE gap is non-monotonic in $\Delta$, with the highest improvement achieved at intermediate noise values.

\begin{figure}[t!]
    \centering
    \includegraphics[width=\linewidth]{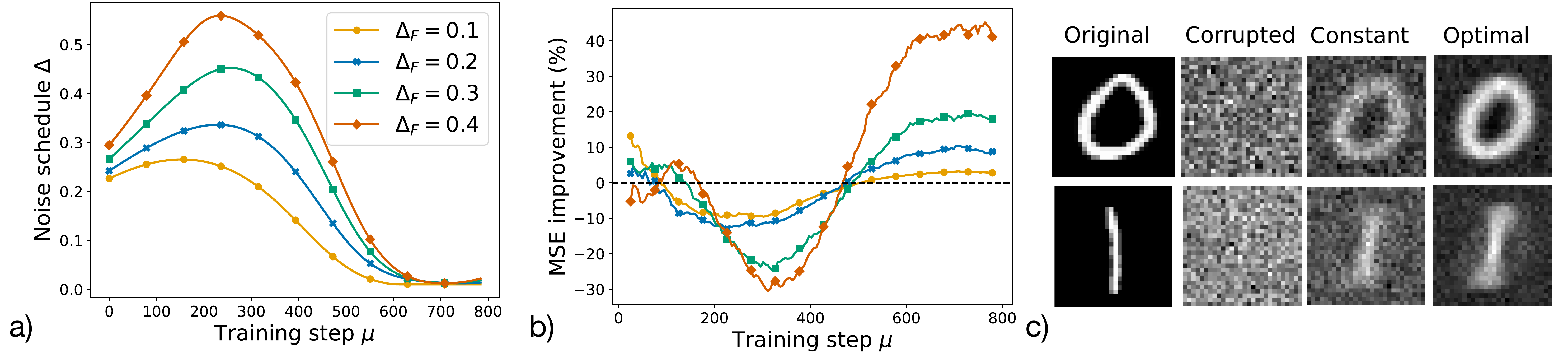}
    \caption{\textbf{Optimal denoising schedules for the $0-1$ MNIST dataset.} {\bf a)} Optimal noise schedule $\Delta$ as a function of the training. \textbf{b)} MSE improvement of the optimal strategy compared to the constant one at $\Delta=\Delta_F$. Each curve is averaged over $10$ random realizations of the training set. \textbf{c)} Examples of images for $\Delta_F=0.4$: original, corrupted, denoised with the constant schedule $\Delta=\Delta_F$, and denoised with the optimal schedule. \textbf{Parameters:} $K=C_1=2$, $\alpha_F=1$, $\eta=\eta_b=5$, $\sigma=0.1$, $N=784$, $g(z)=z$. \textbf{Initialization:} $b=0$. Other initial conditions and parameters are given in the Supplementary Material.}
    \label{fig:mnist_dae}
\end{figure}

We next demonstrate the applicability of our framework to real-world data by focusing on the MNIST dataset, which consists of labeled \(28 \times 28\) grayscale images of handwritten digits from $0$ to $9$. For simplicity, we restrict our analysis to the digits $0$ and $1$. We numerically estimate the mean vectors ${\bm \mu}_{1,1}$ and ${\bm \mu}_{1,2}$, corresponding to the digit classes $0$ and $1$, respectively, as well as the standard deviations $\sigma_{1,1}$ and $\sigma_{1,2}$. 
Considering learning trajectories with $\alpha_F = 1$, we identify the optimal noise schedule $\Delta$ for different values of the testing noise $\Delta_F$. The results are shown in Fig.~\ref{fig:mnist_dae}, and all exhibit a characteristic pattern: an initial increase in noise followed by a gradual decrease toward the end of the training trajectory. As expected, higher values of the testing noise $\Delta_F$ lead to overall higher noise levels throughout the schedule.

We then use these schedules to train a DAE with $K=2$ on a randomly selected training set of $P=784$ images (corresponding to $\alpha_F = P/N = 1$). In Fig.~\ref{fig:mnist_dae}b, we observe that the optimal noise schedule yields MSE improvements of up to $40\%$ relative to the constant strategy $\Delta = \Delta_F$. This improvement is also apparent in the denoised images shown in Fig.~\ref{fig:mnist_dae}c. These results highlight the practical benefits of optimizing the noise schedule, confirming the applicability of our theoretical framework to real data.

\section{Discussion}
\blue{
We have introduced a systematic framework that combines statistical physics and control theory to identify optimal training protocols in a broad range of learning scenarios. We have illustrated the applicability of this approach through several examples spanning hyperparameter tuning, architecture design, and data selection. We have consistently found that optimal training protocols outperform standard heuristics and can exhibit highly nontrivial structures that would be difficult to guess a priori.}

\blue{In curriculum learning, we have shown that non-monotonic difficulty schedules can outperform both easy-to-hard and hard-to-easy curricula. The optimal protocol balances two competing objectives, maximizing teacher-student alignment along relevant directions and suppressing the growth of irrelevant weight components, and achieves this balance through a structured easy-hard-easy schedule. Crucially, the joint optimization of the curriculum and the learning rate reveals a synergy between these two controls: a properly adjusted learning rate renders an easy-to-hard curriculum globally optimal, and the optimal schedule naturally incorporates a warm-up phase followed by a sharp learning rate drop at the curriculum transition.}

\blue{In dropout-regularized networks, the optimal schedule delays the onset of regularization, exploiting the early training phase to maximize signal alignment before suppressing harmful co-adaptations among hidden nodes. This delayed-dropout strategy outperforms constant dropout and reproduces, via a principled derivation, heuristic annealed schedules that had previously been proposed on empirical grounds. The order parameters of the theory provide a transparent post-hoc interpretation: the transition from full to reduced dropout coincides with the saturation of the teacher-student cosine similarity, making the switching criterion explicit and interpretable.}

\blue{For denoising autoencoders, the optimal noise schedule typically features an initial decrease followed by a moderate increase toward the end of training. This nontrivial structure simultaneously enhances the reconstruction capability of the autoencoder and accelerates convergence of the skip connection toward its optimal value. The optimal batch augmentation schedule progressively increases the batch size throughout training, effectively annealing gradient variance and stabilizing learning in the later stages. }

\blue{
The order parameters of the theory systematically reveal interpretable structures underlying the optimal strategies, thereby connecting protocol design to fundamental learning trade-offs. Our approach complements and extends earlier work on optimal learning rates in committee machines \cite{rattray1998analysis,saad1997globally,schlosser1999optimization} and continual learning \cite{mori2024optimal}, providing a unified treatment that accommodates a broader range of controls, objectives, and constraints.}

\blue{\section{Limitations and perspectives}
The results presented in this work are obtained within a class of models actively studied in current theoretical machine learning and statistical physics literature, where analytical tractability is key to derive asymptotically sharp predictions. However, this setting relies on several idealizing assumptions, each of which represents an opportunity for future investigation. As detailed in Section~\ref{sec:special_cases}, the current formulation already accommodates a variety of learning settings beyond those investigated here, including dynamic architectural features such as gating and attention. An important direction for future work is to understand whether and how structure in real data qualitatively changes the optimal schedules. A feasible next step is to study this question within our framework using tractable synthetic models that encode data structure \cite{goldt2020hmm,adomaityte2023classification,PhysRevLett.131.027301}. It would also be relevant to analyze batch learning settings allowing to study how training schedules affect the interplay between memorization and generalization, e.g., via dynamical mean-field theory \cite{mignacco2020dynamical}. Additionally, it would be relevant to extend the analysis to deep and overparametrized architectures.
Furthermore, the discussion in Section~\ref{sec:DAE} on optimal noise schedules could be extended to generative settings such as diffusion models, enabling the derivation of optimal noise injection protocols. Alternative training objectives could also be considered, including fairness metrics, robustness under distribution shift, or computational efficiency. While we analyzed gradient-based learning rules, it would be interesting to explore biologically-plausible update mechanisms. Finally, it is of practical importance to extend our optimal-control framework to state-of-the-art ML models, where order parameters are not clearly defined. A pragmatic route could be to learn low-dimensional collective variables via clustering and projection, an approach that has proven effective in other contexts, such as agent-based opinion dynamics \cite{monti2025hierarchical}.
}

\paragraph{Software availability.} All code required to reproduce the figures presented in this work is publicly available at \url{https://github.com/francescomori/optimal_learning/}.

\paragraph{Acknowledgements.} We thank S.~Sarao Mannelli and A.~Sclocchi for helpful discussions. We are grateful to H.~Cui for useful feedback on the manuscript. 

\paragraph{Funding.} F.~Mignacco was supported by the Simons Foundation (Award Number: 1141576) and the Human Frontier Science Program (Award Number: LT0049/2023-L). F.~Mori was supported by a Leverhulme Trust International Professorship grant (Award Number: LIP-2020-014), the Center of Mathematical Sciences and Applications (CMSA) of Harvard University, and a Schuman Educational Research Grant.

\bibliography{apssamp}
\bibliographystyle{unsrt}

\clearpage
\appendix
\renewcommand{\thesection}{S\arabic{section}}
\renewcommand{\thesubsection}{S\arabic{section}.\arabic{subsection}}
\renewcommand{\thefigure}{S\arabic{figure}}
\renewcommand{\theequation}{S\arabic{equation}}
\renewcommand{\thetable}{S\arabic{table}}
\setcounter{section}{0}
\setcounter{figure}{0}
\setcounter{equation}{0}
\setcounter{table}{0}

\begin{center}
{\Large\bfseries Supplementary Material}\\[0.5em]
{\large A statistical physics framework for optimal learning}\\[0.5em]
Francesca Mignacco, Francesco Mori
\end{center}

\section{Derivation of the learning dynamics}
\label{appendix:dynamics}
In this section, we derive the set of ordinary differential equations (ODEs) for the order parameters given the main text, that track the dynamics of online stochastic gradient descent (SGD). We consider the cost function
\begin{equation}
\label{eq:cost_appendix}
    \mathcal{L}({\bm w},{\bm v}|\bm{x},\bm{c})=\ell\left(\frac{{\bm x}^{\top} {\bm w_*}}{\sqrt{N}},\frac{{\bm x}^{\top}{\bm w}}{\sqrt{N}},\frac{\ww^\top\ww}{N} ,{\bm v}, {\bm c},z\right) + \tilde{g}\left(\frac{\ww^\top\ww}{N},{\bm v}\right)\,.
\end{equation}
The update rules for the network's parameters are
\begin{align}
\label{eq:updaterules_supmat}
\begin{split}
 \ww^{\mu+1}=\ww^\mu-\eta\nabla_{\ww}\mathcal{L}(\ww^\mu,\vv^\mu|\bm{x}^\mu,\bm{c}^\mu)=\ww^\mu -\eta \left[\frac{{\xx^\mu} \nabla_2\ell^\mu}{\sqrt{N}} +2\frac{\ww^\mu \nabla_3\ell^\mu}{N}+2\frac{\ww^\mu\nabla_1 \tilde g^\mu}{N}\right]\;,
 \end{split}\\
     \begin{split}
        \vv^{\mu+1}=\vv^\mu -\frac{\eta}{N}\nabla_4\ell^\mu-\frac{\eta}{N}\nabla_2\tilde g^\mu\;,
    \end{split}
\end{align}
where we use $\nabla_k \ell$ to denote the gradient of the function $\ell$ with respect to its $k^{\rm th}$ argument, with the convention that it is reshaped as a matrix of the same dimensions of that argument, e.g., $\nabla_2\ell\in\mathbb{R}^{L\times K}$. For simplicity, we omit the function's arguments, by only keeping the time dependence, i.e., $\ell^\mu=\ell\left(\frac{{{\bm x}^\mu}^{\top} {\bm w_*}}{\sqrt{N}},\frac{{{\bm x}^\mu}^{\top}{\bm w}^\mu}{\sqrt{N}},\frac{{\ww^\mu}^\top\ww^\mu}{N} ,{\bm v}^\mu, {\bm c}^\mu,z^{\mu}\right)$. For a given realization of the cluster coefficients $\cc$, we introduce the compact notation $\mmu_{\cc}\in\mathbb{R}^{N\times L}$ to denote the matrix with columns $\mmu_{l,c_l}$. It is useful to define the local fields  
\begin{align}
\llambda^\mu=\frac{{\xx^\mu}^{\top}\ww^\mu}{\sqrt{N}}\in\mathbb{R}^{L\times K}\;, && \llambda_*^\mu=\frac{{\xx^\mu}^{\top}\ww_*}{\sqrt{N}}\in\mathbb{R}^{L\times M}\;, &&\rrho^\mu_{\bm c}=\frac{{\xx^\mu}^{\top} \mmu_{\cc}}{\sqrt N}\in\mathbb{R}^{L\times L}\;.
\end{align}
Notice that, due to the online-learning setup, at each training step the input $\xx$ is independent of the
weights. Therefore, due to the Gaussianity of the inputs, the local fields are also jointly Gaussian with
zero mean and second moments given by: 
\begin{align}
\begin{split}
\mathbb{E}_{\xx|\cc}\left[\lambda_{lk}\lambda_{l'k'}\right]&=\frac{{\ww_k}\cdot\mmu_{l,c_l}}{{N}}\frac{{\ww_{k'}}\cdot\mmu_{l',c_{l'}}}{{N}}+\delta_{l,l'}\,\sigma^2_{l,c_l}\frac{{\ww_k}\cdot \ww_{k'}}{N}\\
&=R_{k(l,c_l)}R_{k'(l',c_{l'})}+\delta_{l,l'}\sigma^2_{l,c_l}Q_{kk'}\;,
\end{split}
\\
\begin{split}
\mathbb{E}_{\xx|\cc}\left[\lambda_{lk}\lambda_{*,l'm}\right]&
=\frac{{\ww_k}\cdot\mmu_{l,c_l}}{{N}}\frac{{\ww_{*,m}}\cdot\mmu_{l',c_{l'}}}{{N}}+\delta_{l,l'}\,\sigma^2_{l,c_l}\frac{{\ww_k}\cdot \ww_{*,m}}{N}\\
&=R_{k(l,c_l)}S_{m(l',c_{l'})}+\delta_{l,l'}\sigma^2_{l,c_l}M_{km}\;,
\end{split}
\\
\begin{split}
\mathbb{E}_{\xx|\cc}\left[\lambda_{*,lm}\lambda_{*,l'm'}\right]
&=\frac{{\ww_{*,m}}\cdot\mmu_{l,c_l}}{{N}}\frac{{\ww_{*,m'}}\cdot\mmu_{l',c_{l'}}}{{N}}+\delta_{l,l'}\,\sigma^2_{l,c_l}\frac{{\ww_{*,m}}\cdot \ww_{*,m'}}{N}\\
&=S_{m(l,c_l)}S_{m'(l',c_{l'})}+\delta_{l,l'}\sigma^2_{l,c_l}T_{mm'}\;,
\end{split}
\end{align}
\begin{align}
\begin{split}
\mathbb{E}_{\xx|\cc}\left[\lambda_{lk}\rho_{{\bm c'}, l'l''}\right]&=\frac{{\ww_{k}}\cdot\mmu_{l,c_l}}{{N}}\frac{\mmu_{l',c_{l'}}\cdot\mmu_{l'',c'_{l''}}}{N}+\delta_{l,l'}\,\sigma^2_{l,c_l}\frac{{\ww_{k}}\cdot \mmu_{l'',c'_{l''}}}{ N}\\
&=R_{k(l,c_l)}\Omega_{(l',c_{l'})(l'',c'_{l''})}+\delta_{l,l'}\sigma^2_{l,c_l}R_{k(l'',c'_{l''})}\;,
\end{split}\\
\begin{split}
\mathbb{E}_{\xx|\cc}\left[\lambda_{*,lm}\rho_{{\bm c}',l'l''}\right]&=\frac{{\ww_{*,m}}\cdot\mmu_{l,c_l}}{{N}}\frac{\mmu_{l',c_{l'}}\cdot\mmu_{l'',c'_{l''}}}N+\delta_{l,l'}\,\sigma^2_{l,c_l}\frac{{\ww_{*,m}}\cdot \mmu_{l'',c'_{l''}}}{ N}\\
&=S_{m(l,c_l)}\Omega_{(l',c_{l'})(l'',c'_{l''})}+\delta_{l,l'}\sigma^2_{l,c_l}S_{m(l'',c'_{l''})}\;,
\end{split}\\
\begin{split}
\mathbb{E}_{\xx|\cc}\left[\rho_{{\bm c}',ll'}\rho_{{\bm c}'',l''l'''}\right]&=\frac{{\mmu_{l',c'_{l'}}}\cdot\mmu_{l,c_l}}N\frac{\mmu_{l'',c_{l''}}\cdot\mmu_{l''',c''_{l'''}}}N+\delta_{l,l''}\,\sigma^2_{l,c_l}\frac{{\mmu_{l',c'_{l'}}}\cdot \mmu_{l''',c''_{l'''}}}N\\
&=\Omega_{(l,c_{l})(l',c'_{l'})}\Omega_{(l'',c_{l''})(l''',c''_{l'''})}+\delta_{l,l''}\sigma^2_{l,c_l}\Omega_{(l',c'_{l'})(l''',c''_{l'''})}\;,
\end{split}
\end{align}
where we have introduced the order parameters
\begin{align}
\begin{split}
\label{eq:orderparams_supmat}
&Q_{kk'}\coloneqq \frac{{\ww_k}\cdot \ww_{k'}}{N}\;, \quad M_{km}\coloneqq\frac{{\ww^\mu_k}\cdot \ww_{*,m}}{N}\;,\quad R_{k(l,c_l)}\coloneqq\frac{{\ww_k}\cdot\mmu_{l,c_l}}{{N}}\;,\\
& S_{m(l,c_l)}\coloneqq\frac{{\ww_{*,m}}\cdot\mmu_{l,c_l}}{{N}}\;,\quad
T_{mm'}\coloneqq \frac{{\ww_{*,m}}\cdot \ww_{*,m'}}{N}\;,\quad \Omega_{(l,c_l)(l',c'_{l'})}=\frac{\mmu_{l,c_l}\cdot\mmu_{l',c'_{l'}}}N\;.
\end{split}
\end{align}
Note that in the expressions above the variable $\bm x$ is drawn with cluster membership $\bm c$ fixed. The additional cluster membership variables, e.g., $\bm c'$ and $\bm c''$ are fixed and do not intervene in the generative process of $\bm x$. The cost function defined in \eqref{eq:cost_appendix} depends on the weights $\ww$ only through the local fields and the order parameters. Similarly, the generalization error can be computed as an average over the local fields 
\begin{align}
\varepsilon_g(\ww,\vv)=\mathbb{E}_{\cc}\mathbb{E}_{(\llambda,\llambda_*)|\cc}\left[\ell_g\left(\llambda_*,\llambda,\bm{Q},\vv,\cc,0\right)\right]\;,
\end{align}
where the function $\ell_g$ may coincide with the loss $\ell$ or denote a different metric depending on the context.

Since the local fields are Gaussian, their distribution is completely specified by the first two moments, which are functions of the order parameters. By substituting the update rules of \eqref{eq:updaterules_supmat} into the definitions in \eqref{eq:orderparams_supmat}, we obtain the following evolution equations governing the order‐parameter dynamics
\begin{align}
\begin{split}\label{eq:supmat_evol_Q}
    &\QQ^{\mu+1}-\QQ^\mu=\frac{{\ww^{\mu+1}}^\top \ww^{\mu+1}}{N}-\frac{{\ww^{\mu}}^\top \ww^{\mu}}{N}=\\&\quad-\frac\eta N \left[{\llambda^\mu}^\top\nabla_2\ell^\mu+\nabla_2{\ell^\mu}^\top{\llambda^\mu}+2\QQ^\mu\left( \nabla_3\ell^\mu+\nabla_1 \tilde g^\mu\right)+2\left({\nabla_3\ell^\mu}+\nabla_1 \tilde g^\mu\right)^\top \QQ^\mu\right]\\
    &\quad +\frac{\eta^2}{N}\left[{\nabla_2\ell^\mu}^\top\frac{{\xx^\mu}^\top{\xx^{\mu}}}{N}\nabla_2\ell^\mu+\mathcal{O}\left(\frac 1N\right)\right]\;,
\end{split}\\
 \begin{split}
        \MM^{\mu+1}-\MM^\mu=\frac{{\ww^{\mu+1}}^\top\ww_*}{N}-\frac{{\ww^{\mu}}^\top\ww_*}{N}=-\frac{\eta}{N}\left[{\nabla_2\ell^\mu}^\top\llambda_*^\mu+2\left(\nabla_3\ell^\mu+\nabla_1\tilde g^\mu\right)^\top\MM^\mu\right]\;,
    \end{split}\\
    \begin{split}
        \RR_{\bm c'}^{\mu+1}-\RR_{\bm c'}^\mu=\frac{{\ww^{\mu+1}}^\top\mmu_{\bm c'}}{{N}}-\frac{{\ww^{\mu}}^\top\mmu_{\bm c '}}{{N}}=-\frac{\eta}{N}\left[{\nabla_2\ell^\mu}^\top{\rrho}_{\bm c'}+2\left(\nabla_3\ell^\mu+\nabla_1\tilde g\right)^\top\RR_{\bm c'}^\mu\right]\;,
    \end{split}
\end{align}
where we have omitted subleading terms in $N$. Note that, while for convenience we write $\RR_{\bm c'}$ for an arbitrary cluster membership variable ${\bm c}'=(c'_1\,,\ldots\,,c'_L)$, it is sufficient to keep track of the scalar variables $R_k,(l,c''_l)$ for $k=1\,,\ldots K$, $l=1\,,\ldots\,,L$, $c''_l=1\,,\ldots\,, C_l$, resulting in $K(C_1+C_2+\ldots+C_L)$ variables. We define a ``training time'' $\alpha=\mu/N$ and take the infinite-dimensional limit $N\rightarrow\infty$ while keeping $\alpha$ of order one. We obtain the following ODEs
\begin{align}
\begin{split} \label{eq:supmat_evol_Q_cont}
    \frac{\dd\QQ}{\dd \alpha}&=\mathbb{E}_{\cc}\Big[-\eta\left\{\mathbb{E}_{\llambda,\llambda_*|\cc}\left[\llambda^\top\nabla_2\ell\right] +2 \,\QQ \left(\mathbb{E}_{\llambda,\llambda_*|\cc}\left[\nabla_3\ell\right]+\nabla_1\tilde g\right) + {\rm (transpose)} \right\}\\
   &\qquad \qquad +\eta^2 \,\mathbb{E}_{\llambda,\llambda_*|\cc}\left[\nabla_2\ell^\top{\rm diag}(\bm{\sigma^2}_{\bm c})\nabla_2\ell\right]\Big]\coloneqq f_{\QQ}\;,
\end{split}\\
\begin{split} \label{eq:supmat_evol_M_cont}
    \frac{\dd \MM}{\dd \alpha}=\mathbb{E}_{\cc}\Big[-\eta \,\mathbb{E}_{\llambda,\llambda_*|\cc}\left[{\nabla_2\ell}^\top\llambda_*\right]-2\eta \left(\mathbb{E}_{\llambda,\llambda_*|\cc}\left[{\nabla_3\ell}\right]+\nabla_1\tilde g\right)^\top\MM\Big]\coloneqq f_{\MM}\;,
\end{split}\\
    \begin{split}
        \frac{\dd\RR_{\bm c'}}{\dd\alpha}=\mathbb{E}_{\cc}\Big[-\eta \,\mathbb{E}_{\llambda,\llambda_*|\cc}\left[\nabla_2\ell^\top\rrho_{\bm c'}\right]-2\eta \left(\mathbb{E}_{\llambda,\llambda_*|\cc}\left[\nabla_3\ell\right]+\nabla_1\tilde g \right)^\top\RR_{\bm c'}\Big]\coloneqq f_{\RR_{\bm c'}}\;,\label{eq:supmat_evol_R_cont}
        \end{split}
\end{align}
where we remind that $\ell=\ell\left(\llambda_*,\llambda,\QQ,\vv,\cc,z\right)$ and $\tilde g =\tilde g (\QQ,\vv)$, and we have defined the vector of variances $\bm{\sigma^2}_{\bm c}=(\sigma^2_{1,c_1},\ldots,\sigma^2_{L,c_L})$. In going from \eqref{eq:supmat_evol_Q} to \eqref{eq:supmat_evol_Q_cont}, we have used 
\begin{equation}
   \lim_{N\to\infty} \frac{\xx_l \cdot \xx_{l'}}{N}= \sigma_{l,c_l}^2\delta_{ll'}\,.
\end{equation}
Crucially, when taking the thermodynamic limit $N\to\infty$, we have replaced the right-hand sides in Eqs.~(\ref{eq:supmat_evol_Q_cont}-\ref{eq:supmat_evol_R_cont}) with their expected value over the data distribution. Indeed, it can be shown rigorously that, under additional assumptions, the fluctuations of the order parameters can be neglected \cite{goldt2019dynamics}. Although we do not provide a rigorous proof of this result here, we verify this concentration property with numerical simulations, see Section \ref{app:sims}. Finally, the additional parameters $\vv$ evolve according to the low-dimensional equations
\begin{align}
    \frac{\dd \vv}{\dd \alpha}=\mathbb{E}_{\cc}\Big[-\eta\,\mathbb{E}_{\llambda,\llambda_*|\cc} \left[\nabla_4\ell +\nabla_2\tilde g\right]\Big]\coloneqq f_{\vv}\;. \label{eq:supmat_evol_v_cont}
\end{align}
To conclude, note that the expectations in Eqs.~(\ref{eq:supmat_evol_Q_cont}), (\ref{eq:supmat_evol_R_cont}) and (\ref{eq:supmat_evol_v_cont}) decompose into an average over the low‐dimensional cluster vector $\mathbf c$, whose distribution is given by the model, and an average over the Gaussian fields $\bm\lambda$ and $\bm\lambda_{*}$, whose moments are fully specified by the order parameters, resulting in a closed-form system of equations. The expectations can be evaluated either analytically or via Monte Carlo sampling.

\subsection{Curriculum learning}
\label{app:cl}

The equations for the curriculum learning problem can be derived as a special case of those of \cite{NEURIPS2022_84bad835}. The misclassification error can be expressed in terms of the order parameters as
\begin{align}
\epsilon_g(\mathbb{Q})=\frac 12 -\frac{1}{\pi}\sin^{-1}\left(\frac{M_{11}}{\sqrt{T(Q_{11}+\Delta Q_{22})}}\right)\;.
\end{align}
The evolution equations for the order parameters can be obtained from \eqref{eq:supmat_evol_Q_cont} and \eqref{eq:supmat_evol_R_cont}, yielding
\begin{align}
    \begin{split}\label{eq:app_ODE_CL}
        \frac{\dd Q_{11}}{\dd \alpha}&=-\bar{\lambda} Q_{11}+\frac{4\eta}{\pi(Q_{11}+\Delta Q_{22}+2)}\left[\frac{M_{11}(\Delta Q_{22}+2)}{\sqrt{T(Q_{11}+\Delta Q_{22}+2)-M_{11}^2}}-\frac{Q_{11}}{\sqrt{Q_{11} +\Delta Q_{22}+1}}\right]\\
        &\qquad+\frac{2}{\pi^2}\frac{\eta^2}{\sqrt{Q_{11}+\Delta Q_{22}+1}}\left[\frac{\pi}{2}+\sin^{-1}\left(\frac{Q_{11}+\Delta Q_{22}}{2+3(Q_{11}+\Delta Q_{22})}\right)\right.\\
        &\qquad \left.-2\sin^{-1}\left(\frac{M_{11}}{\sqrt{\left(3(Q_{11}+\Delta Q_{22})+2\right)}\sqrt{T(Q_{11}+\Delta Q_{22}+1)-M_{11}^2 }}\right)\right]
        \,,\\
        \frac{\dd Q_{22}}{\dd \alpha}&=-\bar{\lambda} Q_{22}-\frac{4\eta\Delta Q_{22}}{\pi(Q_{11}+\Delta Q_{22}+2)}\left[\frac{M_{11}}{\sqrt{T(Q_{11}+\Delta Q_{22}+2)-M_{11}^2}}+\frac{1}{\sqrt{Q_{11}+\Delta Q_{22}+1}}\right]\\
        &\qquad +\frac{2}{\pi^2}\frac{\Delta \eta^2}{\sqrt{Q_{11}+\Delta Q_{22}+1}}\left[\frac{\pi}{2}+\sin^{-1}\left(\frac{Q_{11}+\Delta Q_{22}}{2+3(Q_{11}+\Delta Q_{22})}\right)\right.\\
        &\qquad\left.-2\sin^{-1}\left(\frac{M_{11}}{\sqrt{\left(3(Q_{11}+\Delta Q_{22})+2\right)}\sqrt{T(Q_{11}+\Delta Q_{22}+1)-M_{11}^2}}\right)\right]\,,\\
        \frac{\dd M_{11}}{\dd \alpha}&=-\frac{\bar{\lambda}}{2}M_{11}+\frac{2\eta}{\pi(Q_{11}+\Delta Q_{22}+2)}\left[\sqrt{T(Q_{11}+\Delta Q_{22}+2)-M_{11}^2}-\frac{M_{11}}{\sqrt{Q_{11}+\Delta Q_{22}+1}}\right]\,,
    \end{split}
\end{align}
where $\bar{\lambda}=\lambda\eta$.

\subsection{Dropout regularization}
\label{app:dropout}

In this section, we provide the expressions of the ODEs and the generalization error for the model of dropout regularization presented in the main text. This model corresponds to $L=C_1=1$, $\bm{\mu}_{1,1}=\bm{0}$, and $\sigma_{1,1}=1$. The derivation of these results can be found in \cite{mori2025analytic}. The generalization error reads
\begin{align}
\begin{split}
    \epsilon_g &= \mathbb{E}_{\bm x}\left[ \frac12 \left(f^*_{\bm w_*}(\bm x)-f^{\rm test}_{\bm w}(\bm x)\right)^2 \right]=\frac{p_f^2}{\pi} \sum_{i,k=1}^K  \arcsin\left(\frac{Q_{ik}}{\sqrt{1 + Q_{ii}} \sqrt{1 + Q_{kk}}}\right)
\\&\quad+ \frac{1}{\pi} \sum_{n,m=1}^K  \arcsin\left(\frac{T_{nm}}{\sqrt{1 + T_{nn}} \sqrt{1 + T_{mm}}}\right)- \frac{2p_f}{\pi} \sum_{i=1}^K \sum_{n=1}^M \arcsin\left(\frac{M_{in}}{\sqrt{1 + Q_{ii}} \sqrt{1 + T_{nn}}}\right).
\end{split}
\end{align}

The ODEs read
\begin{align}
\frac{\mathrm{d} M_{in}}{\mathrm{d} \alpha} = f_{M_{in}}(Q,M),  &&
\frac{\mathrm{d} Q_{ik}}{\mathrm{d} \alpha} = f_{Q_{ik}}(Q,M),  
\end{align}
Introducing the notation
\begin{equation}
    \mathcal{N}\left[p,\{i,j,k,\ldots ,l\}\right] = p^n\,,
\end{equation}
where \( n = |\{i,j,k,\ldots ,l\}| \) is the cardinality of the set \(\{i,j,k,\ldots ,l\}\), we find 
\begin{align} \label{eq:app_dropout_M}
f_{M_{in}}&\equiv \eta  \left[ 
    \sum_{m=1}^M \mathcal{N}\left[p,\{i\}\right] I_3(i, n, m) - \sum_{j=1}^K  \mathcal{N}\left[p,\{i,j\}\right]  I_3(i, n, j)
\right], \\
f_{Q_{ik}}&\equiv \eta  \left[ 
    \sum_{m=1}^M \mathcal{N}\left[p,\{i\}\right]  I_3(i, k, m) - \sum_{j=1}^K \mathcal{N}\left[p,\{i,j\}\right] I_3(i, k, j)
\right] \notag \\
&\quad + \eta \left[ 
    \sum_{m=1}^M \mathcal{N}\left[p,\{k\}\right] I_3(k, i, m) - \sum_{j=1}^K \mathcal{N}\left[p,\{k,j\}\right] I_3(k, i, j)
\right] \notag \\
&\quad + \eta^2  \Bigg[ 
    \sum_{n=1}^M \sum_{m=1}^M \mathcal{N}\left[p,\{i,k\}\right]  I_4(i, k, n, m)  \notag  - 
    2 \sum_{j=1}^K \sum_{n=1}^M \mathcal{N}\left[p,\{i,k,j\}\right]  I_4(i, k, j, n) \notag\\
&\quad\quad + \sum_{j=1}^K \sum_{l=1}^K \mathcal{N}\left[p,\{i,j,k,l\}\right] I_4(i, k, j, l) + \mathcal{N}\left[p,\{i,k\}\right] \sigma^2 J_2(i, k)
\Bigg], \label{eq:app_dropout_Q}
\end{align}
where 
\begin{align}
J_2 &\equiv  \frac{2}{\pi} \left( 1 + c_{11} + c_{22} + c_{11} c_{22} - c_{12}^2 \right)^{-1/2}, \\
I_2 &\equiv \frac{1}{\pi} \arcsin\left(\frac{c_{12}}{\sqrt{1 + c_{11}} \sqrt{1 + c_{12}}}\right),  \\
I_3 &\equiv \frac{2}{\pi} \frac{1}{\sqrt{\Lambda_3}} \frac{c_{23}(1 + c_{11}) - c_{12} c_{13}}{1 + c_{11}}, \\
I_4 &\equiv  \frac{4}{\pi^2} \frac{1}{\sqrt{\Lambda_4}} \arcsin\left(\frac{\Lambda_0}{\sqrt{\Lambda_1 \Lambda_2}}\right), 
\end{align}
and
\begin{align}
\Lambda_4 &= (1 + c_{11})(1 + c_{22}) - c_{12}^2, \\
\Lambda_3 &=(1+c_{11})*(1+c_{33})-c_{13}^2\,,\\
\Lambda_0 &= \Lambda_4 c_{34} - c_{23} c_{24}(1 + c_{11}) - c_{13} c_{14}(1 + c_{22}) +   c_{12} c_{13} c_{24}+c_{12} c_{14} c_{23} , \\
\Lambda_1 &= \Lambda_4 (1 + c_{33}) - c_{23}^2(1 + c_{11}) - c_{13}^2(1 + c_{22}) + 2 c_{12} c_{13} c_{23},\\
\Lambda_2 &= \Lambda_4 (1 + c_{44}) - c_{24}^2(1 + c_{11}) - c_{14}^2(1 + c_{22}) + 2 c_{12} c_{14} c_{24}.
\end{align}
The indices $i,j,k,l$ and $n,m$ indicate the student's and the teacher's nodes, respectively. For compactness, we adopt the notation for $I_2$, $I_3$, and $I_4$ of Ref.~\cite{goldt2019dynamics}. As an example, $I(i,n)$ takes as input the correlation matrix of the preactivations corresponding to the indices $i$ and $n$, i.e., $\lambda_i={\bm w}_i\cdot {\bm x}/\sqrt{N}$ and $\lambda_{*,n}={\bm w}^*_n \cdot {\bm x}/\sqrt{N}$. For this example, the correlation matrix would be
\begin{equation}
    C=\begin{pmatrix}
c_{11} & c_{12} \\
c_{21} & c_{22}
\end{pmatrix}=\begin{pmatrix}
\langle\lambda_i\lambda_i \rangle & \langle \lambda_i \lambda_{*,n}\rangle \\
\langle \lambda_{*,n} \lambda_i\rangle& \langle \lambda_{*,n}\lambda_{*,n}\rangle\end{pmatrix}=\begin{pmatrix}
Q_{ii} & M_{in} \\
M_{in} & T_{nn}
\end{pmatrix}
\,.
\end{equation}

\subsection{Denoising autoencoder}
\label{appendix:dynamics_dae}

We define the additional local fields
\begin{align}
    \tilde{\lambda}_k\equiv \frac{{ \tilde{\bm x}}\cdot{\bm w}_k}{\sqrt{N}}=\sqrt{1-\Delta} \lambda_{1,k}+\sqrt{\Delta}\lambda_{2,k}\,, \quad \tilde{\rho}_{{\bm c}, l }\equiv \frac{{ \tilde{\bm x}} \cdot{\bm \mu}_{l,c_l}}{\sqrt{N}}=\sqrt{1-\Delta} \rho_{{\bm c}, 1l}+\sqrt{\Delta}\rho_{{\bm c},2l}\,,
\end{align}
where we recall $\lambda_{1,k}= {\bm w}_k\cdot {\bm x}_1/\sqrt{N}$, $ \lambda_{2,k}= {\bm w}_k\cdot {\bm x}_2/\sqrt{N}$, $\rho_{{\bm c},1l}= {\bm \mu}_{l,c_l}\cdot {\bm x}_1/\sqrt{N}$, $ \rho_{{\bm c},2l}= {\bm \mu}_{l,c_l}\cdot {\bm x}_2/\sqrt{N}$. Here, we take $C_2=1$ and $\mmu_{2,c_2}={\bm 0}$, so that $\rho_{{\bm c},12}=\rho_{{\bm c},22}=\tilde{\rho}_{{\bm c},2}=0$. The local fields are Gaussian variables with moments given by 
\begin{align}
    \mathbb{E}_{\bm x|\bm c} \left[\lambda_{1,k}\right] = \frac{{\bm w}_k\cdot {\bm \mu}_{1,c_1}}{N}= R_{k(1,c_1)}\,,&& \mathbb{E}_{\bm x|\bm c} \left[  \rho_{{\bm c'},11}\right]= \frac{\mmu_{1,c_1}\cdot\mmu_{1,c'_1}}{N}= \Omega_{(1,c_1)(1,c'_1)}\,,
\end{align}
\begin{align}
 \mathbb{E}_{\bm x|\bm c} \left[ \lambda_{2,k}\right]&=   \mathbb{E}_{\bm x|\bm c} \left[  \rho_{{\bm c'},2l}\right]=0\;,
 \end{align}
 \begin{align}\mathbb{E}_{\bm x|\bm c} \left[ \lambda_{1,k} \lambda_{2,h}\right] =\mathbb{E}_{\bm x|\bm c} \left[ \lambda_{1,k} \rho_{{\bm c'},2l}\right]=\mathbb{E}_{\bm x|\bm c} \left[ \lambda_{2,k} \rho_{{\bm c'},1l}\right]=\mathbb{E}_{\bm x|\bm c} \left[ \rho_{{\bm c'},1l} \rho_{{\bm c'},2l'}\right]= 0\,,
\end{align}
\begin{align}
    \mathbb{E}_{\bm x|\bm c} \left[  \lambda_{1,k}\lambda_{1,h}\right]=R_{k(1,c_1)}R_{h(1,c1)}+\sigma^2_{1,c_1} Q_{kh}\,,&&
    \mathbb{E}_{\bm x|\bm c} \left[  \lambda_{2,k}\lambda_{2,h}\right]= Q_{kh}\,,
\end{align}
\begin{align}
    \mathbb{E}_{\bm x|\bm c} \left[  \tilde\lambda_{j}\lambda_{1,k}\right]=\sqrt{1-\Delta}\mathbb{E}_{\bm x|\bm c} \left[  \lambda_{1,k}\lambda_{1,j}\right]\,,&&
    \mathbb{E}_{\bm x|\bm c} \left[  \tilde\lambda_{j}\lambda_{2,k}\right]=\sqrt{\Delta}\mathbb{E}_{\bm x|\bm c} \left[  \lambda_{2,k}\lambda_{2,j}\right]\,,
\end{align}
\begin{align}
    \mathbb{E}_{\bm x|\bm c} \left[  \rho_{{\bm c}',11}^2\right]=\Omega_{(1,c_1)(1,c'_1)}^2+\sigma^2_{1,c_1} \Omega_{(1,c'_1)(1,c_1)}\,,&&
    \mathbb{E}_{\bm x|\bm c} \left[  \rho_{{\bm c'},21}^2\right]= \Omega_{(1,c'_1)(1,c'_1)}\,.
\end{align}
\begin{align}
    \mathbb{E}_{\bm x|\bm c} \left[  \lambda_{1,k}\rho_{{\bm c'},11}\right] =\sigma_{1,c_1}^2 R_{k(1,c'_1)}+\Omega_{(1,c'_1)(1,c_1)} R_{k(1,c_1)}\,,&&  \mathbb{E}_{\bm x|\bm c} \left[  \lambda_{2,k}\rho_{{\bm c'},21}\right] = R_{k(1,c'_1)}\,.
\end{align}
It is also useful to compute the first moments of the combined variables 
\begin{align}
\mathbb{E}_{\bm x|\bm c} \left[  \tilde \lambda_k\right]=\sqrt{1-\Delta}\, R_{k(1,c_1)}\;, && \mathbb{E}_{\bm x|\bm c} \left[  \tilde \rho_{{\bm c'},1}\right]=\sqrt{1-\Delta}\, \Omega_{(1,c_1)(1,c'_1)}\,,
\end{align}
and the second moments
\begin{align}
\begin{split}
    \mathbb{E}_{\bm x|\bm c} \left[ \tilde \lambda_k \tilde \lambda_h\right]-\mathbb{E}_{\bm x|\bm c} \left[  \tilde \lambda_k\right] \mathbb{E}_{\bm x|\bm c} \left[  \tilde \lambda_h\right]&=\left[(1-\Delta)\sigma_{1,c_1}^2 +\Delta\right]Q_{kh}\,,\\
    \mathbb{E}_{\bm x|\bm c} \left[ \tilde \rho_{{\bm c'},1}^2\right]-\mathbb{E}_{\bm x|\bm c} \left[  \tilde \rho_{{\bm c'},1}\right] ^2&=\left[(1-\Delta)\sigma_{1,c_1}^2 +\Delta\right]\Omega_{(1,c'_1)(1,c'_1)}\,.
\end{split}
\end{align}
Finally, we have
\begin{align}
    \mathbb{E}_{\bm x|\bm c} \left[  \tilde \lambda_{k}\rho_{{\bm c'},11}\right]&=\sqrt{1-\Delta}\mathbb{E}_{\bm x|\bm c} \left[  \lambda_{1,k}\rho_{{\bm c'},11}\right]\,,\\ \mathbb{E}_{\bm x|\bm c} \left[  \tilde\lambda_{k}\tilde \rho_{{\bm c'},1}\right]&=(1-\Delta) \mathbb{E}_{\bm x|\bm c} \left[  \lambda_{1,k}\rho_{{\bm c'},11}\right]+\Delta \mathbb{E}_{\bm x|\bm c} \left[  \lambda_{2,k}\rho_{{\bm c'},21}\right]\,.
\end{align}
The mean squared error (MSE) can be expressed in terms of the order parameter as follows
\begin{align}
    \begin{split}
    \text{MSE}(\ww,b)&= \mathbb{E}_{\bm{x},\bm{c}}\left[\Vert \xx-f_{\ww,b}(\tilde{\xx}) \Vert_2^2\right]=\mathbb{E}_{\bm{c}}\left\{N\left[\sigma_k^2\left(1-b\sqrt{1-\Delta}\right)^2+b^2\Delta\right]\right.\\&\quad+\left.\sum_{j,k=1}^K Q_{jk} \mathbb{E}_{\bm{x}|\bm{c}}\left[ g(\tilde\lambda_j) g(\tilde\lambda_k)\right]-2\sum_{k=1}^K\mathbb{E}_{\bm{x}|\bm{c}}\left[ (\lambda_{1k}-b\tilde\lambda_k)g(\tilde\lambda_k)\right] \right\}\,,
\end{split}
\end{align}
where we have neglected constant terms. The weights are updated according to
\begin{align}
\begin{split}\ww^{\mu+1}_k&=\ww^{\mu}_k+\frac{\eta}{\sqrt{N}}g \left(\tilde \lambda^{\mu}_k\right)\left(\xx_1^{\mu}-b\,\tilde{\xx}^{\mu}-\sum_{h=1}^K \frac{{\ww_h^{\mu}}}{\sqrt{N}}g\left(\tilde\lambda^{\mu}_h\right)\right)\\&\quad+\frac{\eta}{\sqrt{N}}g'(\tilde \lambda^{\mu}_k)\,\left(\lambda^{\mu}_{1,k}-b\,\tilde{\lambda}^{\mu}_k-\sum_{h=1}^K\frac{\ww^{\mu}_k\cdot{\ww^{\mu}_h}}{{N}}g\left(\tilde\lambda^{\mu}_h\right)\right)\,{\tilde{\xx}^{\mu}} \;,
    \end{split}
\end{align}
The skip connection is also trained with SGD. To leading order, we find
\begin{align}
b^{\mu+1}=b^{\mu}+\frac{\eta_b}{N}\left(\sqrt{1-\Delta}\sigma_{1,c_1}^2-b^{\mu}(1-\Delta)\sigma_{1,c_1}^2-b^{\mu}\Delta\right)\;.
\end{align}
Note that, conditioning on a given cluster $c_1$, for large $N$, we have 
\begin{align}
   \frac 1 N {\xx_1  \cdot \xx_1} \underset{N\gg 1}{\approx}\sigma_{1,c_1}^2 \,, \quad \frac 1 N {\tilde{\xx} \cdot \tilde{\xx}} \underset{N\gg 1}{\approx}(1-\Delta)\sigma_{1,c_1}^2 + \Delta\,,\quad \frac 1 N {\xx_1 \cdot \tilde{\xx}} \underset{N\gg 1}{\approx}\sqrt{1-\Delta}\,\sigma_{1,c_1}^2 \,.
\end{align}
For simplicity, we will consider the linear activation $g(z)=z$. In this case, it is possible to derive explicit equations for the evolution of the order parameters as follows:
\begin{align}
\begin{split}R^{\mu+1}_{k (1,c'_1)}&=R^{\mu}_{k (1,c'_1)}+\frac{\eta}{{N}}\mathbb{E}_{\bm c}\left[\mathbb{E}_{\bm x|\bm c}\left[ \tilde\lambda^{\mu}_k\rho^{\mu}_{\bm{c'},11}\right]-2b\mathbb{E}_{\bm x|\bm c}\left[\tilde\lambda^{\mu}_k\tilde\rho^{\mu}_{\bm{c'},1}\right]-\sum_{j=1}^K R^{\mu}_{j (1,c'_1)} \mathbb{E}_{\bm x|\bm c} \left[\tilde\lambda^{\mu}_k \tilde\lambda^{\mu}_j\right]\right.\\
    &\quad +\left.\mathbb{E}_{\bm x|\bm c}\left[\lambda^{\mu}_{1,k}\tilde\rho^{\mu}_{\bm{c'},1}\right]-\sum_{j=1}^K Q_{jk}\mathbb{E}_{\bm x|\bm c}\left[\tilde\lambda^{\mu}_j\tilde\rho_{\bm{c'},1}\right]\right]\;,\label{eq:app_DAE_ODE_r}
    \end{split}
\end{align} 
\begin{align}
    \begin{split}
Q^{\mu+1}_{jk}&=Q^{\mu}_{jk}+\frac{\eta}{N}\mathbb{E}_{\bm c}\left\{\left(\mathbb{E}_{\bm x|\bm c}\left[\tilde\lambda_j\Lambda_k\right]+\mathbb{E}_{\bm x|\bm c}\left[\tilde\lambda_k\Lambda_j\right]\right)\left[2+\eta\left(\frac{\xx_1 \cdot \tilde{\xx}}{N}-b\frac{\tilde{\xx} \cdot \tilde{\xx}}{N}\right)\right]+\eta\mathbb{E}_{\bm x|\bm c}\left[\Lambda_j\Lambda_k\right]\frac{\tilde{\xx}\cdot\tilde{\xx}}{N}\right.\\
        &\quad +\left.\eta \mathbb{E}_{\bm x|\bm c}\left[\tilde\lambda_j\tilde \lambda_k\right]\left(\frac{\xx_1\cdot \xx_1}{N}-2b \frac{\xx_1\cdot \tilde{\xx}}{N}+b^2\frac{\tilde{\xx}\cdot\tilde{\xx}}{N}\right)\right\}\\
        &=Q^{\mu}_{jk}+\frac{\eta}{N}\mathbb{E}_{\bm c}\left\{\left(\mathbb{E}_{\bm x|\bm c}\left[\tilde\lambda_j\Lambda_k\right]+\mathbb{E}_{\bm x|\bm c}\left[\tilde\lambda_k\Lambda_j\right]\right)\left[2+\eta\left(\sqrt{1-\Delta}\sigma_{1,c_1}^2-b ((1-\Delta)\sigma_{1,c_1}^2 + \Delta))\right)\right]\right.\\
        & +\eta\mathbb{E}_{\bm x|\bm c}\left[\Lambda_j\Lambda_k\right]((1-\Delta)\sigma_{1,c_1}^2 + \Delta) +\left.\eta \mathbb{E}_{\bm x|\bm c}\left[\tilde\lambda_j\tilde \lambda_k\right]\left(\sigma_{1,c_1}^2-2b \sqrt{1-\Delta}\,\sigma_{1,c_1}^2+b^2((1-\Delta)\sigma_{1,c_1}^2 + \Delta))\right)\right\}
    \end{split}
\end{align}
where we have introduced the definition
\begin{align}
\label{eq:defLLambda}
    \Lambda_k \equiv \lambda_{1,k} -b \tilde\lambda_k -\sum_{j=1}^K Q_{jk}\tilde\lambda_j\;.
\end{align}
We can compute the averages
\begin{align}
    \begin{split}
\mathbb{E}_{\bm x|\bm c}\left[\tilde\lambda_j\Lambda_k\right]&=\mathbb{E}_{\bm x|\bm c}\left[\tilde\lambda_j\lambda_{1,k}\right] -\sum_{i=1}^K\left(b\delta_{ik}+Q_{ki}\right)\mathbb{E}_{\bm x|\bm c}\left[\tilde\lambda_j\tilde\lambda_i\right]\;,\\ \mathbb{E}_{\bm x|\bm c}\left[\Lambda_j\Lambda_k\right]&=\mathbb{E}_{\bm x|\bm c}\left[\lambda_{1,j}\lambda_{1,k}\right]-\sum_{i=1}^K\left(b\delta_{ij}+Q_{ji}\right)\mathbb{E}_{\bm x|\bm c}\left[\tilde\lambda_i\lambda_{1,k}\right]-\sum_{i=1}^K\left(b\delta_{ik}+Q_{ki}\right)\mathbb{E}_{\bm x|\bm c}\left[\tilde\lambda_i\lambda_{1,j}\right]\\
        &\quad +\sum_{i,\ell=1}^K \left(b\delta_{ik}+Q_{ki}\right)\left(b\delta_{\ell j}+Q_{j\ell}\right)\mathbb{E}_{\bm x|\bm c}\left[\tilde\lambda_i\tilde\lambda_\ell\right]\;.
    \end{split}
\end{align}
Finally, it is useful to evaluate the MSE in the special case of linear activation:
\begin{align}
\begin{split}
\text{MSE}&=\mathbb{E}_{\bm c}\left\{N\left[\sigma_{1,c_1}^2\left(1-b\sqrt{1-\Delta}\right)^2+b^2\Delta\right]\right.\\
&+\sum_{j,k=1}^K Q_{jk}\left[\left((1-\Delta)\sigma_{1,c_1}^2+\Delta\right)Q_{jk}+(1-\Delta)R_{j,(1,c_1)}R_{k,(1,c_1)}\right]\\&-2\left.\sum_{k=1}^K\left[\sqrt{1-\Delta}\sigma_{1,c_1}^2 Q_{kk}-b\left[\left((1-\Delta)\sigma_{1,c_1}^2+\Delta\right)Q_{kk}+(1-\Delta)R_{k,(1,c_1)}^2\right]\right]\right\}\;.
\end{split}
\end{align}

\subsubsection{Data augmentation}

 We consider inputs $\bm{x}=(\bm{x}_1, \bm{x}_{2}, \ldots, \bm{x}_{B+1})\in\mathbb{R}^{N\times B+1}$, where $\xx_1\sim\mathcal{N}\left(\frac{\mmu_{1,c_1}}{\sqrt N},\sigma^2\bm{I}_N\right)$ denotes the clean input and $\bm{x}_{2}, \ldots, \bm{x}_{B+1}\overset{\rm i.i.d.}{\sim}\mathcal{N}(\bm{0},\bm{I}_N)$. Each clean input $\xx_1$ is used to create multiple corrupted samples: $\tilde{\xx}_a=\sqrt{1-\Delta}\,\xx_1+\sqrt{\Delta}\,\xx_{a+1}$, $a=1,\ldots,B$, that are used as a mini-batch for training. The SGD dynamics of the tied weights modifies as follows:
\begin{align}
\begin{split}\ww^{\mu+1}_k=\\\ww^{\mu}_k+\frac{\eta}{B^{\mu}\sqrt{N}}\sum_{a=1}^{B^\mu}\left\{\tilde \lambda^{\mu}_{a,k}\left(\xx_1^\mu-b\,\tilde{\xx}^{\mu}_a-\sum_{j=1}^K\frac{{\ww^{\mu}_j}}{\sqrt{N}}\tilde\lambda^{\mu}_{a,j}\right)+\left(\lambda^{\mu}_{1,k}-b\,\tilde{\lambda}^{\mu}_{a,k}-\sum_{j=1}^K\frac{\ww^{\mu}_k\cdot{\ww^{\mu}_j}}{{N}}\tilde\lambda^{\mu}_{a,j}\right)\,{\tilde{\xx}^\mu}_a\right\}\;,
    \end{split}
\end{align}
where 
\begin{equation}
\tilde{\lambda}_{a,k}=\frac{\bm{\tilde x}_a\cdot \bm{w}_k}{\sqrt{N}}=\sqrt{1-\Delta}\lambda_{1,k}+\sqrt{\Delta}\lambda_{a+1,k}\,.
\end{equation}
While the equations for $b$ and $M$ remain unchanged, we need to include additional terms in the equation for $Q$. We find
\begin{align}
    \begin{split}
Q^{\mu+1}_{jk}&=Q^{\mu}_{jk}+\frac{\eta}{N}\mathbb{E}_{\bm c}\left\{\left(\mathbb{E}_{\bm x|\bm c}\left[\tilde\lambda_j\Lambda_k\right]+\mathbb{E}_{\bm x|\bm c}\left[\tilde\lambda_k\Lambda_j\right]\right)\left[2+\frac{\eta}{B}\left(\sqrt{1-\Delta}\sigma_{1,c_1}^2-b ((1-\Delta)\sigma_{1,c_1}^2 + \Delta))\right)\right]\right.\\
        & +\frac{\eta}{B}\mathbb{E}_{\bm x|\bm c}\left[\Lambda_j\Lambda_k\right]((1-\Delta)\sigma_{1,c_1}^2 + \Delta) \\ &+\frac{\eta}{B} \mathbb{E}_{\bm x|\bm c}\left[\tilde\lambda_j\tilde \lambda_k\right]\left(\sigma_{1,c_1}^2-2b \sqrt{1-\Delta}\,\sigma_{1,c_1}^2+b^2((1-\Delta)\sigma_{1,c_1}^2 + \Delta))\right)\\
    &+\frac{\eta(B-1)}{B}(1-\Delta)\mathbb{E}_{\bm x|\bm c}\left[\Lambda_{a,j}\Lambda_{a',k}\right]\sigma_{1,c_1}^2\\
    &+\frac{\eta(B-1)}{B}\mathbb{E}_{\bm x|\bm c}\left[\tilde\lambda_{a,j}\tilde\lambda_{a',k}\right]\left(\left(1+b^2(1-\Delta)\right)\sigma_{1,c_1}^2-2b\sqrt{1-\Delta}\sigma_{1,c_1}^2\right)\\
        & +\left.\frac{\eta(B-1)}{B }\left(\mathbb{E}_{\bm x|\bm c}\left[\tilde\lambda_{a,j}\Lambda_{a',k}\right]+\mathbb{E}_{\bm x|\bm c}\left[\tilde\lambda_{a,k}\Lambda_{a',j}\right]\right)\left(\sqrt{1-\Delta}\sigma_{1,c_1}^2 -b (1-\Delta)\sigma_{1,c_1}^2\right)
 \right\}\label{eq:app_DAE_Q_augm}
    \end{split}
\end{align}
We derive the following expressions for the average quantities, valid for $a\neq a'$
\begin{align}
    \mathbb{E}_{\bm x|\bm c} \left[\tilde\lambda_{a,j}  \tilde\lambda_{a',k}\right] &= (1-\Delta) \mathbb{E}_{\bm x|\bm c} \left[ \lambda_{1,j} \lambda_{1,k}\right]\,,\\
       \mathbb{E}_{\bm x|\bm c} \left[ \tilde\lambda_{a,j}  \Lambda_{a',k}\right] &= \left[\sqrt{1-\Delta}-b(1-\Delta)\right]\mathbb{E}_{\bm x|\bm c} \left[ \lambda_{1,j} \lambda_{1,k}\right] -(1-\Delta)\sum_{i=1}^K Q_{ki}\mathbb{E}_{\bm x|\bm c} \left[ \lambda_{1,j}\lambda_{1,i}\right]\,,\\
       \mathbb{E}_{\bm x|\bm c} \left[ \Lambda_{a,j}  \Lambda_{a',k}\right] &=(1-b\sqrt{1-\Delta})^2\mathbb{E}_{\bm x|\bm c} \left[ \lambda_{1,j}\lambda_{1,k}\right]
       +(1-\Delta) \sum_{i,h=1}^K Q_{ji}Q_{kh}\mathbb{E}_{\bm x|\bm c} \left[ \lambda_{1,i}\lambda_{1,h}\right]\\
       +&\left[b(1-\Delta)-\sqrt{1-\Delta}\right]\sum_{i=1}^K \left(Q_{ji}\mathbb{E}_{\bm x|\bm c} \left[ \lambda_{1,k}\lambda_{1,i}\right]+Q_{ki}\mathbb{E}_{\bm x|\bm c} \left[ \lambda_{1,j}\lambda_{1,i}\right]\right)\,,\nonumber
\end{align}
where $\Lambda_{a,j}$ is defined as in \eqref{eq:defLLambda}.

\section{Optimal control methods}
\label{appendix:oc_methods}
\subsection{Indirect methods} 
\label{sec:oc_indirect_methods}

We consider the cost functional
\begin{equation}
    \mathcal{F}[\bm{u}]=\epsilon_g(\mathbb{Q}(\alpha_F))\,,\label{eq:functional_generalization}
\end{equation}
where the square brackets indicate functional dependence on the full control trajectory $\bm{u}(\alpha)$, for $0\leq\alpha\leq\alpha_F$. Following Pontryagin’s maximum principle \cite{pontryagin1957some}, we introduce the Lagrange multipliers $\hat{\mathbb{Q}}(\alpha)$ to enforce the training dynamics 
\begin{equation}
    \frac{\dd\mathbb{Q}(\alpha)}{\dd\alpha}
        =f_{\mathbb{Q}}\bigl(\mathbb{Q}(\alpha),\,\bm{u}(\alpha)\bigr)\,.
\end{equation}
We obtain
\begin{equation}
    \mathcal{F}[\bm{u},\mathbb{Q},\hat{\mathbb{Q}}]
    = \epsilon_g\bigl(\mathbb{Q}(\alpha_F)\bigr)
    + \int_{0}^{\alpha_F} {\rm d}\alpha\;\hat{\mathbb{Q}}(\alpha)\cdot
      \left[
        -\frac{\dd\mathbb{Q}(\alpha)}{\dd\alpha}
        + f_{\mathbb{Q}}\bigl(\mathbb{Q}(\alpha),\,\bm{u}(\alpha)\bigr)
      \right],\label{eq:augmented_F}
\end{equation}
where $\hat{\mathbb{Q}}(\alpha)$ are known as adjoint (or costate) variables. The optimality conditions are
    $\delta \mathcal{F}/\delta \hat{\mathbb{Q}}(\alpha) = 0$ and $\delta \mathcal{F}/\delta \mathbb{Q}(\alpha) = 0$.
The first yields the forward dynamics \eqref{eq:ODE_compact}. For $\alpha<\alpha_F$, the second, after integration by parts, gives the adjoint (backward) ODEs
\begin{align}
    -\frac{\dd \hat{\mathbb{Q}}(\alpha)^\top}{\dd \alpha}
    &= \hat{\mathbb{Q}}(\alpha)^\top \nabla_{\mathbb{Q}} f_{\mathbb{Q}}\bigl(\mathbb{Q}(\alpha),\bm u(\alpha)\bigr),
    \label{eq:backward_dyn}
\end{align}
with the final condition at $\alpha=\alpha_F$:
\begin{equation}
    \hat{\mathbb{Q}}(\alpha_F)
    = \nabla_{\mathbb{Q}}\,\epsilon_g\bigl(\mathbb{Q}(\alpha_F)\bigr).
    \label{eq:final_cond}
\end{equation}
Variations at $\alpha=0$ are not considered since $\mathbb{Q}(0)=\mathbb{Q}_0$ is fixed.  
Finally, optimizing $\bm u$ point-wise yields
\begin{equation}
    \bm u^*(\alpha)
    = \underset{\bm u\in\mathcal U}{\arg\min}\;\bigl\{\hat{\mathbb{Q}}(\alpha)\cdot\,f_{\mathbb{Q}}\bigl(\mathbb{Q}(\alpha),\bm u\bigr)\bigr\}.
    \label{eq:optimality_condition}
\end{equation}
In practice, we use the forward-backward sweep method: starting from an initial guess for $\bm u$, we iterate the following steps until convergence.
\begin{enumerate}
  \item Integrate $\mathbb{Q}$ forward via \eqref{eq:ODE_compact} from $\mathbb{Q}(0)=\mathbb{Q}_0$.
  \item Integrate $\hat{\mathbb{Q}}$ backward via \eqref{eq:backward_dyn} from $\hat{\mathbb{Q}}(\alpha_F)$ in \eqref{eq:final_cond}.
  \item Update $\bm u^{k+1}(\alpha)=\gamma_{\rm damp}\bm u^{k}(\alpha)+(1-\gamma_{\rm damp}) \bm u^*(\alpha)$, where $\bm u^*(\alpha)$ is given in \eqref{eq:optimality_condition}.
\end{enumerate}
We typically choose the damping parameter $\gamma_{\rm damp}>0.9$. Convergence is usually reached within a few hundred to a few thousand iterations.

\subsection{Direct methods} 

\label{sec:direct_methods} 

Direct methods discretize the control trajectory $\bm u(\alpha)$ on a finite grid of $I=\alpha_F/\dd\alpha$ intervals and map the continuous‐time OC problem into a finite‐dimensional nonlinear program (NLP). We introduce optimization variables for $\mathbb Q$ and $\bm u$ at each node $\alpha_j=j~\dd\alpha$, enforce the dynamics \eqref{eq:ODE_compact} via constraints on each interval, and solve the resulting NLP using the CasADi package \cite{andersson2019casadi}. In this paper, we implement a multiple‐shooting scheme: $\bm u(\alpha)$ is parameterized as constant on each interval, and continuity of $\mathbb Q$ is enforced at the boundaries. While direct methods are conceptually simpler---relying on standard NLP solvers and avoiding the explicit derivation of adjoint equations---in the settings under consideration, we find that they tend to perform worse when the control $\bm u$ has discrete components. Conversely, indirect methods require computing costate derivatives but yield more accurate solutions for discrete controls. Depending on the problem setting, we therefore choose between direct and indirect approaches as specified in each case. Numerical implementations of both methods are available at \cite{github_repo}. 

\section{Experiments on CIFAR-10}
\label{sec:experiments}

\begin{figure}[t]
    \centering
    \includegraphics[width=\textwidth]{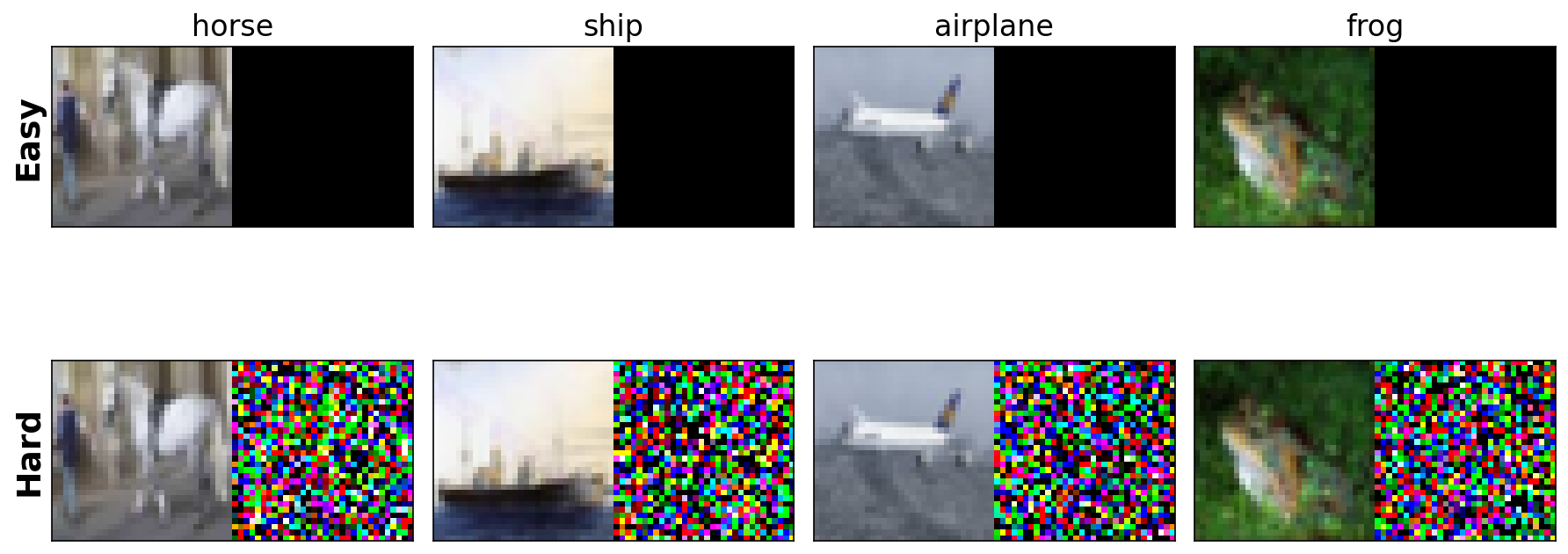}
    \caption{Example composite images from the Cluttered CIFAR-10 dataset.
    The left half of each composite is the target, while the right half is the distractor.
    \textbf{Top row (Easy)}: the distractor is black ($\sigma = 0$).
    \textbf{Bottom row (Hard)}: the distractor is Gaussian noise with $\sigma = 1.0$, making the task more challenging.}
    \label{fig:dataset_examples}
\end{figure}

Following \cite{NEURIPS2022_84bad835}, we construct a Cluttered CIFAR-10 classification task.
Each input is a $3 \times 32 \times 64$ composite image formed by concatenating a $32 \times 32$ CIFAR-10 target image (left half) with a $32 \times 32$ distractor (right half).
For easy samples, the distractor is a black (all-zeros) image.
For \emph{hard} samples, the distractor is drawn from a Gaussian distribution $\mathcal{N}(0, \sigma^2)$ applied independently to each pixel and channel, with $\sigma = 1.0$. All 10 CIFAR-10 classes are used, so that the random-chance test error is $10\%$. Figure~\ref{fig:dataset_examples} shows example composite images.

Motivated by the optimal curriculum derived in the main text, we parameterize the Easy-Hard-Easy training schedule with a single parameter $a \in [0, 50]$, expressed as a percentage of the total training set.
Training proceeds in three phases:
\begin{enumerate}
    \item \textbf{Initial easy phase}: the first $a\%$ of training samples are easy;
    \item \textbf{Hard phase}: the next $50\%$ are hard;
    \item \textbf{Final easy phase}: the remaining $(50 - a)\%$ are easy.
\end{enumerate}
Every schedule presents exactly $50\%$ easy and $50\%$ hard samples, only the presentation order differs.
At the extremes, $a = 0$ corresponds to the Anti-Curriculum strategy, and $a = 50$ corresponds to the standard Curriculum strategy.

The $N = 50{,}000$ CIFAR-10 training images are randomly partitioned into $25{,}000$ easy and $25{,}000$ hard samples. This assignment is held constant across all values of $a$, so that every schedule trains on the \emph{same} set of easy and hard images, only the ordering changes. Each image always receives the same noise realization across all experimental conditions. Within the easy and hard groups, samples are shuffled with a seed that varies per repetition, so that different runs see different within-group orderings while maintaining the same phase structure. The test set consists of the $10{,}000$ CIFAR-10 test images, constructed as a fixed $50/50$ easy/hard mixture.

We use a simple convolutional neural network with three convolutional layers:
\begin{itemize}
    \item Conv2d($3 \to 32$, $3 \times 3$, padding$=1$) $\to$ ReLU $\to$ MaxPool($2 \times 2$)
    \item Conv2d($32 \to 64$, $3 \times 3$, padding$=1$) $\to$ ReLU $\to$ MaxPool($2 \times 2$)
    \item Conv2d($64 \to 64$, $3 \times 3$, padding$=1$) $\to$ ReLU $\to$ MaxPool($2 \times 2$)
\end{itemize}
followed by a fully connected classifier: Flatten $\to$ Linear($64 \times 4 \times 8 \to 256$) $\to$ ReLU $\to$ Linear($256 \to 10$).
The input dimensionality is $3 \times 32 \times 64$ to accommodate the composite images. Training is performed online: each training sample is presented exactly once, in mini-batches of size~$16$.
We use the Adam optimizer with a constant learning rate of $10^{-3}$ and cross-entropy loss.
No learning rate scheduler is employed.

\begin{figure}[t]
    \centering
    \includegraphics[width=0.6\textwidth]{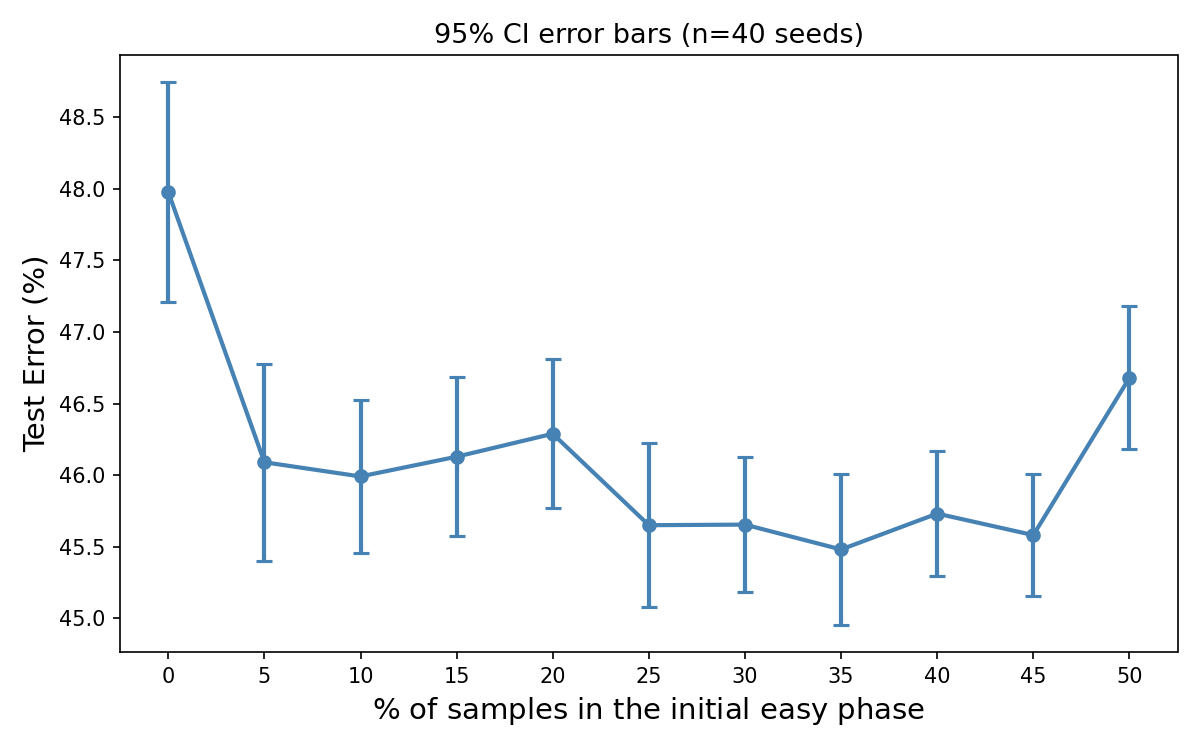}
    \caption{Test error (mean over 30 model initialization seeds, with error bar corresponding to $95\%$ confidence intervals) as a function of the initial easy fraction~$a$ on Cluttered CIFAR-10. The extremes $a = 0$ (Anti-Curriculum) and $a = 50$ (Curriculum) yield the highest error, while intermediate values achieve the lowest error, confirming the advantage of the Easy-Hard-Easy strategy. Random-chance error is \blue{$90\%$}.}
    \label{fig:sweep_a}
\end{figure}

For each value of $a$, we construct the corresponding easy-hard-easy training sequence and train the model from scratch (single online pass). To quantify variability due to model initialization, we repeat each run $30$ times with distinct weight initialization seeds, while keeping the training sequence identical across repetitions. In Figure~\ref{fig:sweep_a}, we show the mean test error as a function of~$a$. Interestingly, we find that the easy-hard-easy, originally derived in the main text for our minimal model, with $a=35\%$ outperforms both Curriculum ($a=50\%)$ and Anti-Curriculum ($a=0\%$).

\section{Supplementary figures and additional details}
\label{appendix:supp_figs}

\begin{figure}
    \centering
\includegraphics[width=0.8\linewidth]{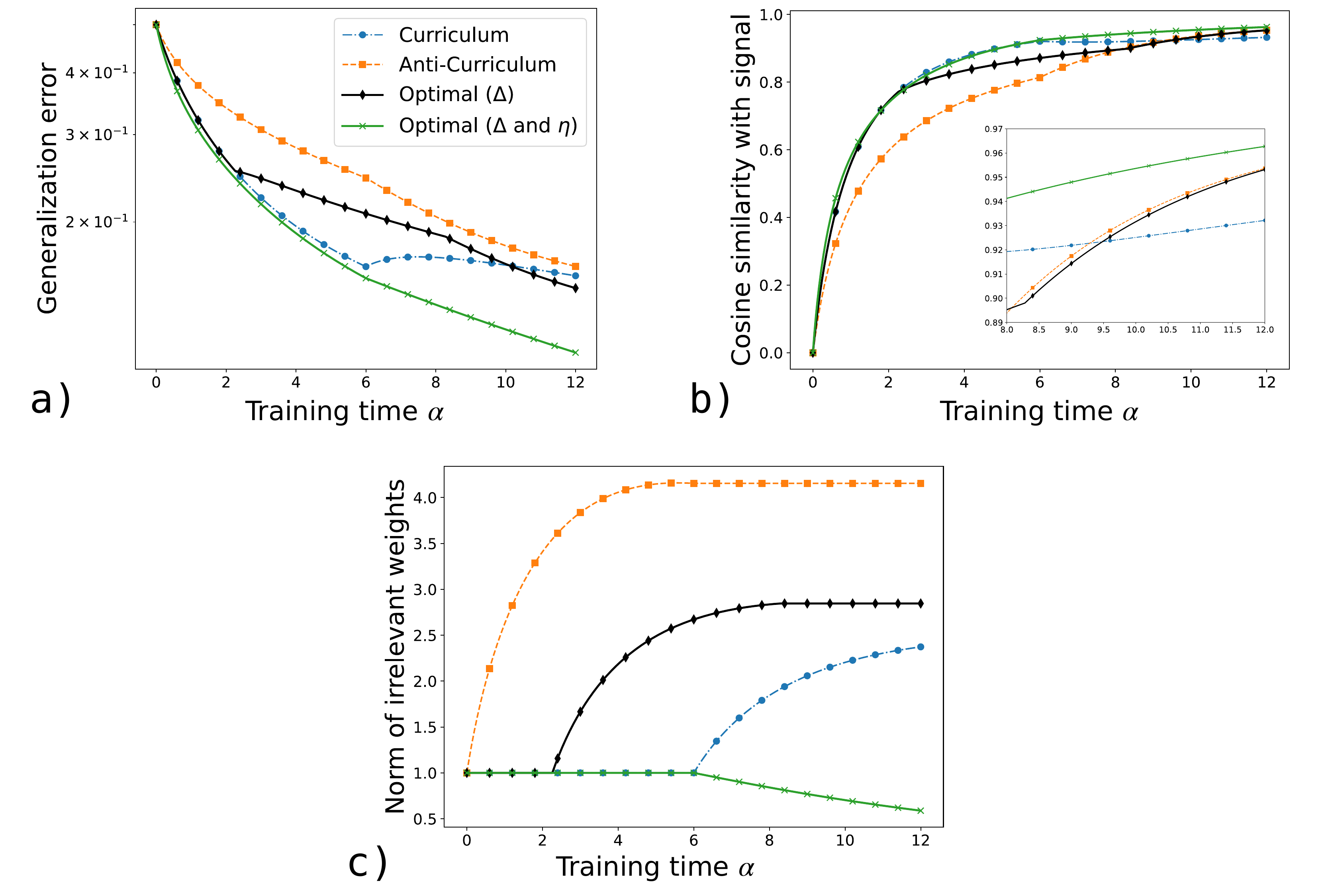}
    \caption{Dynamics of the curriculum learning problem under different training schedules—curriculum (easy to hard) at $\eta=3$, anti-curriculum (hard to easy) at $\eta=3$, the optimal difficulty protocol at $\eta=3$, and the optimal protocol obtained by jointly optimizing $\Delta$ and $\eta$.
    \textbf{(a)} Generalization error vs.~normalized training time $\alpha=\mu/N$. 
    \textbf{(b)} Cosine similarity $M_{11}/\sqrt{T Q_{11}}$ with the target signal (inset zooms into the late-training regime). 
    \textbf{(c)} Squared norm of irrelevant weights $Q_{22}$ vs.\ $\alpha$. {\bf Parameters: }$\alpha_F=12$, $\Delta_1=0$, $\Delta_2=2$, $\eta=3$, $\lambda=0$, $T=2$. {\bf Initial conditions: } $Q_{11}=Q_{22}=1$, $M_{11}=0$.}
    \label{fig:curr_comparison_eta}
\end{figure}

The initial conditions for the order parameters used in Fig.~3 of the main text are
\begin{align}
&\nonumber R = \frac{{\bm w}^{\top}{\bm \mu}_{\cc}}{N}=
\begin{pmatrix}
0.116 & 0.029\\
-0.005 & 0.104
\end{pmatrix}\,,
\qquad Q =\frac{{\bm w}^{\top}{\bm w}}{N}=
\begin{pmatrix}
0.25 & 0.003 \\
0.003 & 0.25
\end{pmatrix}\,,\\
 &\Omega_{(1,1)(1,1)}= \frac{{\bm \mu}_{1,1}\cdot{\bm \mu}_{1,1}}{N}=0.947 \,, \qquad\Omega_{(1,2)(1,2)}= \frac{{\bm \mu}_{1,2}\cdot{\bm \mu}_{1,2}}{N}=0.990\,.
\end{align}
The initial conditions for the order parameters used in Fig.~4 are
\begin{align}
&\nonumber R = \frac{{\bm w}^{\top}{\bm \mu}_{\cc}}{N}=
\begin{pmatrix}
0.339 & 0.200\\
0.173 & 0.263
\end{pmatrix}\,,
\qquad Q =\frac{{\bm w}^{\top}{\bm w}}{N}=
\begin{pmatrix}
1 & 0.00068 \\
0.00068 & 1
\end{pmatrix}\,,\\
 &\Omega_{(1,1)(1,1)}= \frac{{\bm \mu}_{1,1}\cdot{\bm \mu}_{1,1}}{N}=1.737 \,, \qquad\Omega_{(1,2)(1,2)}= \frac{{\bm \mu}_{1,2}\cdot{\bm \mu}_{1,2}}{N}=1.158\,.
\end{align}
The test set used in Fig.~4b contains $13996$ examples. The standard deviations of the clusters are $\sigma_{1,1}=0.05$ and $\sigma_{1,2}=0.033$. The cluster membership probability is $p_c([c_1=1,c_2=1])=0.47$ and $p_c([c_1=2,c_2=1])=0.53$.
The initial conditions for the order parameters used in Fig.~\ref{fig:DAE_sims} are
\begin{align}
&\nonumber R = \frac{{\bm w}^{\top}{\bm \mu}_{\cc}}{N}=
\begin{pmatrix}
0.099 & -0.005\\
-0.002 & 0.102
\end{pmatrix}\,,
\qquad Q =\frac{{\bm w}^{\top}{\bm w}}{N}=
\begin{pmatrix}
0.25 & -0.002 \\
-0.002 & 0.25
\end{pmatrix}\,,\\
 &\Omega_{(1,1)(1,1)}= \frac{{\bm \mu}_{1,1}\cdot{\bm \mu}_{1,1}}{N}=0.976 \,, \qquad\Omega_{(1,2)(1,2)}= \frac{{\bm \mu}_{1,2}\cdot{\bm \mu}_{1,2}}{N}=1.014\,.\label{eq:initial_condition_app_sims}
\end{align}
\blue{These initial conditions are obtained by either randomly generating the centroids ${\bm \mu}_{1,1}$ and ${\bm \mu}_{1,2}$ or by fitting them to the data (MNIST datasets). The student parameters ${\bm w}$ are also randomly generated with a finite overlap with the centroids.}

\section{Numerical simulations}
\label{app:sims}

\begin{figure}
    \centering
    \includegraphics[width=0.8\linewidth]{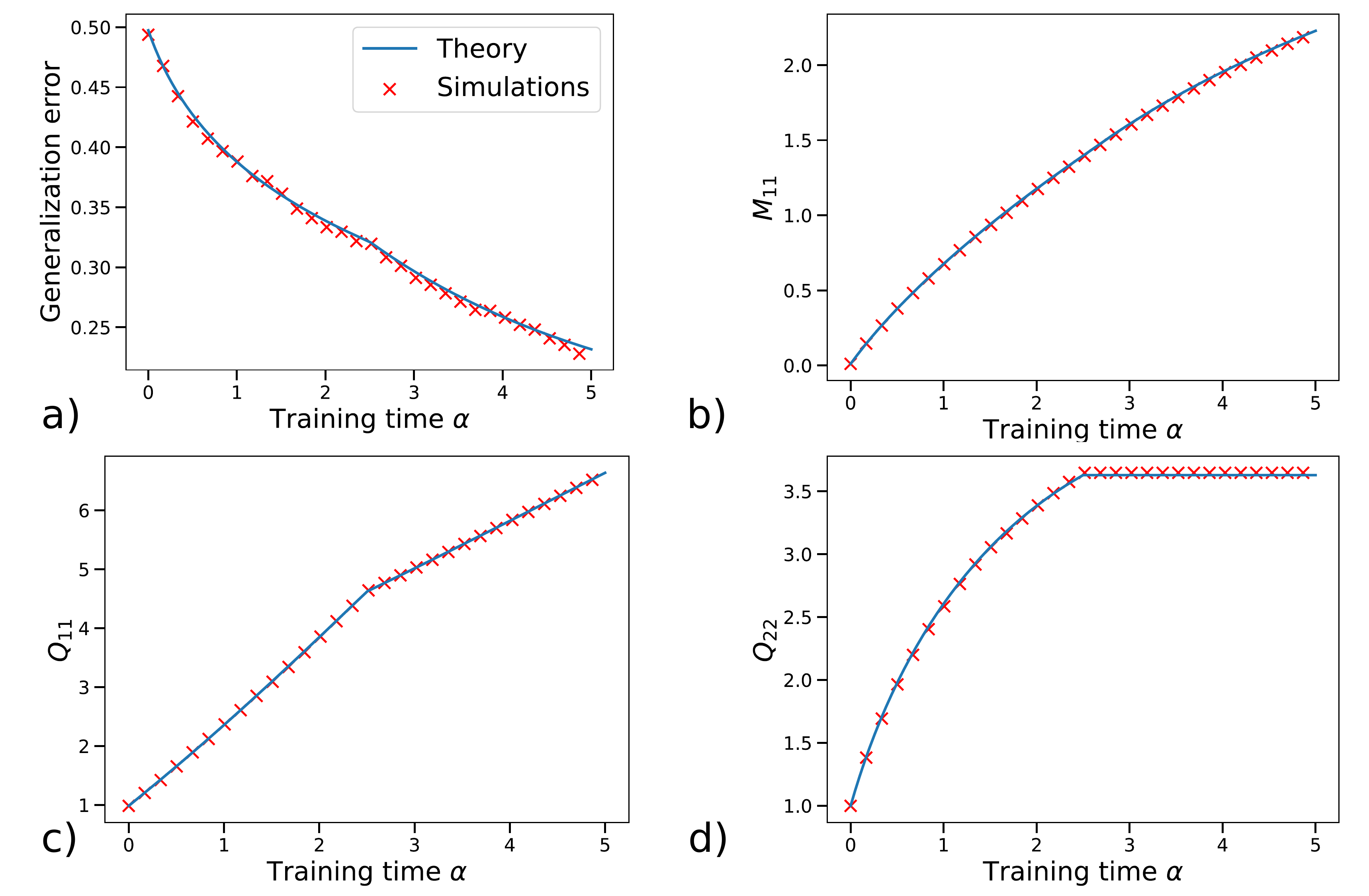}
    \caption{Comparison between theory and simulations in the curriculum learning problem: {\bf a)} generalization error, {\bf b)} teacher-student overlap $M_{11}$, {\bf c)} squared norm $Q_{11}$ of the relevant weights, and {\bf d)} squared norm $Q_{22}$ of the irrelevant weights. The continuous blue lines have been obtained by integrating numerically the ODEs in Eqs.~(\ref{eq:app_ODE_CL}), while the red crosses are the results of numerical simulations of a single trajectory with $N=30000$. The protocol is anti-curriculum with equal proportions of easy and hard samples. {\bf Parameters:} $\alpha_F=5$, $\lambda=0$, $\eta=3$, $\Delta_1=0$, $\Delta_2=2$, $T_{11}=1$. {\bf Initial conditions: $Q_{11}=0.984$, $Q_{22}=0.998$, $M_{11}=0.01$.} 
    \label{fig:CL_sims}
    }
\end{figure}

\begin{figure}
    \centering
    \includegraphics[width=0.8\linewidth]{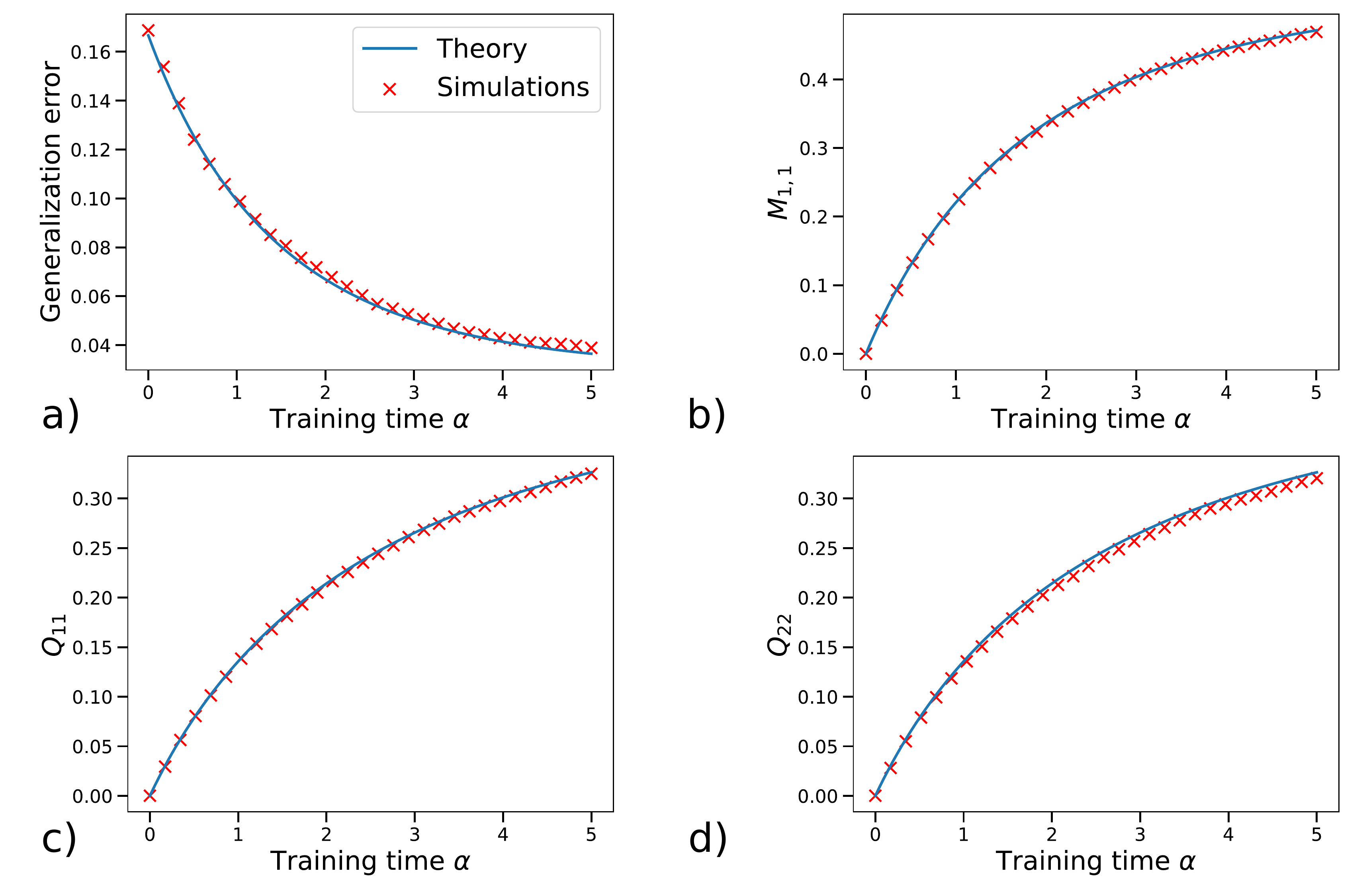}
    \caption{Comparison between theory and simulations for dropout regularization: {\bf a)} generalization error, {\bf b)} teacher-student overlap $M_{1,1}$, {\bf c)} squared norm $Q_{11}$, and {\bf d)} squared norm $Q_{22}$. The continuous blue lines have been obtained by integrating numerically the ODEs in Eqs.~(\ref{eq:app_dropout_M}-\ref{eq:app_dropout_Q}), while the red crosses are the results of numerical simulations of a single trajectory with $N=30000$. {\bf Parameters:} $\alpha_F=5$, $\eta=1$, $\sigma_n=0.3$, $p(\alpha)=p_f=0.7$, $T_{11}=1$. {\bf Initial conditions: $Q_{ij}=M_{nk}=0$.} 
    \label{fig:dropout_sims}
    }
\end{figure}

\begin{figure}
    \centering
    \includegraphics[width=0.8\linewidth]{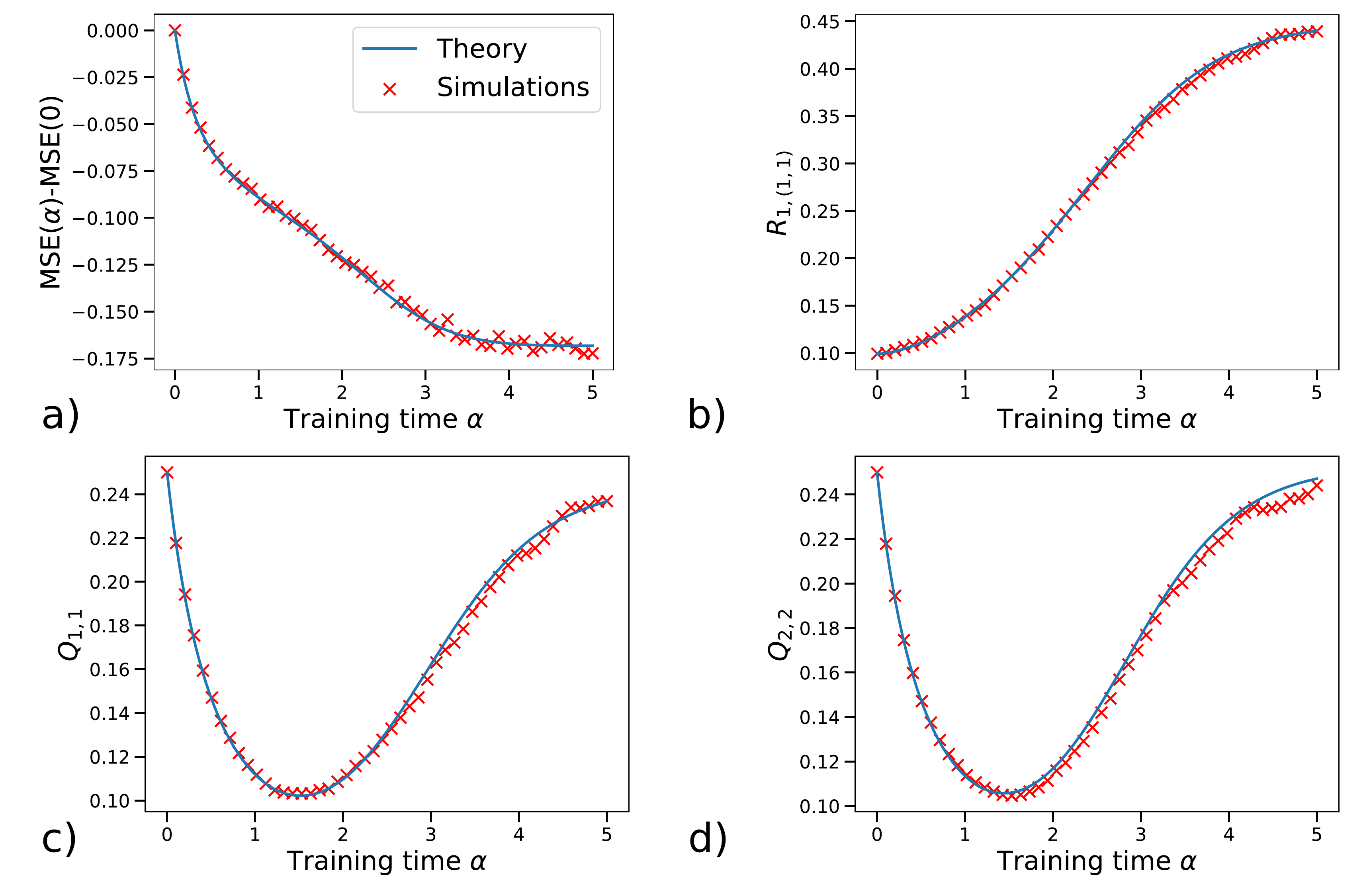}
    \caption{Comparison between theory and simulations for the denoising autoencoder model: {\bf a)} mean square error improvement, {\bf b)} student-centroid overlap $R_{1,(1,1)}$, {\bf c)} squared norm $Q_{11}$. The continuous blue lines have been obtained by integrating numerically the ODEs in Eqs.~(\ref{eq:app_DAE_ODE_r}) and (\ref{eq:app_DAE_Q_augm}), while the red crosses are the results of numerical simulations of a single trajectory with $N=10000$. {\bf Parameters:} $\alpha_F=1$, $\eta=2$, $B(\alpha)=\Bar{B}=5$, $K=C_1=2$, $\sigma=0.1$, $g(z)=z$, $\Delta(\alpha)=\Delta_F=0.8$. The skip connection $b$ is fixed ($\eta_b=0$) to the optimal value $b^*$. Initial conditions are given in \eqref{eq:initial_condition_app_sims}.
    \label{fig:DAE_sims}
    }
\end{figure}

In this section, we validate our theoretical predictions against numerical simulations for the three scenarios studied: curriculum learning (Fig.~\ref{fig:CL_sims}), dropout regularization (Fig.~\ref{fig:dropout_sims}), and denoising autoencoders (Fig.~\ref{fig:DAE_sims}). For each case, the theoretical curves are obtained by numerically integrating the respective ODEs, obtained in the high-dimensional limit $N\to\infty$. The simulations are instead obtained for a single SGD trajectory at large but finite $N$. We observe good agreement between theory and simulations. The code for reproducing the numerical simulations is available at \cite{github_repo}.

\end{document}